\documentclass[a4paper,american,times,5p]{elsarticle}

\makeatletter
\def\@journal{ArXiV}
\makeatother

\usepackage[T1]{fontenc}
\usepackage[utf8]{inputenc}
\usepackage{babel}

\usepackage{amsmath}
\usepackage{amsthm}
\usepackage{amssymb}
\usepackage{graphicx}
\usepackage{color}
\usepackage{bm}
\usepackage{booktabs}

\usepackage{stmaryrd}
\usepackage{stackrel}

\makeatletter

\providecommand{\tabularnewline}{\\}

\definecolor{note_fontcolor}{rgb}{1, 0.667969, 0}

\usepackage{environ}
\usepackage[colorinlistoftodos,prependcaption,textsize=small]{todonotes}
\RenewEnviron{lyxgreyedout}{\todo[inline]{\protect\BODY}}

\usepackage{etoolbox}

  \theoremstyle{definition}
  
  \providecommand{\definitionname}{Definition}
  \theoremstyle{plain}
  
  \providecommand{\lemmaname}{Lemma}
  \theoremstyle{remark}
  \newtheorem*{rem*}{\protect\remarkname}
  \providecommand{\remarkname}{Remark}
  \theoremstyle{plain}
  
  \providecommand{\theoremname}{Theorem}

\renewcommand{\vec}[1]{\bm{#1}}

\usepackage{hyperref}

\makeatother

\newcommand{\dd}{\mathrm{d}}

\newcommand{\T}{\mathrm{T}}

\newcommand{\R}{\mathbb{R}}

\newcommand{\J}{\mathcal{J}}

\begin{document}
\begin{frontmatter}

\title{Focusing the Latent Heat Release in 3D Phase Field Simulations of Dendritic Crystal Growth}


\author[fnspe]{Pavel Strachota\corref{cor1}}\ead{pavel.strachota@fjfi.cvut.cz}
\author[fnspe]{Ale\v{s} Wodecki}\ead{ales.wodecki@fjfi.cvut.cz}
\author[fnspe]{Michal Bene\v{s}}\ead{michal.benes@fjfi.cvut.cz}

\cortext[cor1]{Corresponding author. Phone: +420 224 358 563}

\address[fnspe]{Department of Mathematics, Faculty of Nuclear Sciences
and Physical Engineering, Czech Technical University in Prague. Trojanova
13, 120 00 Praha 2, Czech Republic}

\begin{abstract}
Abstract We investigate a family of phase field models for simulating dendritic growth of a pure supercooled substance. The central object of interest is the reaction term in the Allen-Cahn equation, which is responsible for spatial distribution of latent heat release during solidification. In this context, several existing forms of the reaction term are analyzed. Inspired by the known conclusions of matched asymptotic analysis, we propose new variants that are simple enough to allow mathematical and numerical analysis and robust enough to be applicable to solidification under very large supercooling. The resulting models are tested in a number of numerical simulations focusing on mesh-dependence and model parameter settings. Despite the phase interface thickness being relatively large to make numerical computations feasible, the obtained results exhibit a good quantitative agreement with experimental data from rapid solidification of nickel melts.
\end{abstract}

\begin{keyword}
Allen-Cahn equation \sep dendritic crystal growth \sep diffuse phase interface \sep phase field \sep rapid solidification \sep reaction term
\end{keyword}


\end{frontmatter}

\newlength \figwidth
\setlength \figwidth {1.0\columnwidth}



\section{\label{sec:Introduction}Introduction}

Phase field modeling \cite{Provatas_Elder-PF_Book,Boettinger-SolidificationMicro_Overview}
has been a popular and universal tool for solving moving boundary
problems in materials science for more than three decades now. The
theoretical foundations laid out by Allen and Cahn \cite{Allen-Cahn-orig}
were later utilized by Caginalp \cite{Caginalp-Stefan_HeleShaw_PF,Caginalp-Convergence-PF_SIF}
to show the relation between the diffuse interface and sharp interface
models of phase transition phenomena. Anisotropic dendritic crystal
growth in pure supercooled melts \cite{Wheeler-PF-Dendrites,Kobayashi-PF-Dendritic,Kupferman-Num_Study_Morphological_Diagram,Elliott_Gardiner-DoubleObstacle_PF}
and binary alloys \cite{Wheeler-PF-Binary_Alloys} was first simulated
numerically in 2D. Later, single-crystal 3D simulations were performed
by Karma and his collaborators \cite{Karma_Rappel-Num_sim_3D_dendrites}
and the used models evolved further to predict quantitative properties
of crystal growth \cite{Karma_Rappel-Quant_phase_field_modeling,Bragard_Karma-Higly_undercooled_solidif,Hoyt_Karma-Atomistic-continuum-modeling-solidif,Ramirez-Phase-Field_BinaryAlloy}.
More recent studies by many authors have been focusing e.g on multicomponent
alloys \cite{Nestler-PhaseField_Multicomponent}, interaction with
fluid flow \cite{Jeong_Goldenfeld_Dantzig-PF-FEM-3D-Flow}, grain
growth \cite{Suwa-2D-phasefield_grain_growth,Suwa-3D-MPF_grain_growth,3D-Grain_growth_boundary_energy_db},
polycrystalline solidification \cite{Takaki-2D_PF_competitive_growth_polycrystalline},
and efficient numerical methods for very large scale parallel simulations
\cite{Rojas_Takaki_Ohno-PF-LBM-dendrite_convection,Takaki-GPU_PF_LBM-growth_and_motion,Takaki-Solidification_supercomputer}.
There have also been many applications of the phase field approach
to problems beyond the scope of materials science such as \cite{PhaseField-Solving_PDE_on_manifold,PF-NS-analysis-fluid-interaction,PF-fracture-crack-propagation,PF-Asy-Analysis,PF-multi-comp-flow}.

In this paper, we investigate a family of phase field models for simulating
dendritic growth of a pure supercooled substance designed in our previous
work \cite{Benes-Math_comp_aspects_solid} and related to \cite{Kobayashi-PF-Dendritic}.
These models were used in qualitative computational studies in both
2D \cite{Benes-Math_comp_aspects_solid,Benes-Comp_studies_Diff_Interface,Benes-Hilhorst-Weidenfeld}
and 3D \cite{ENUMATH2011} and efficient parallel numerical solvers
were implemented by means of the finite volume method \cite{ALGORITMY2016}.
Extensions to polycrystalline solidification were also implemented
\cite{ISPMA14-Pavel_Ales-ActaPhPoloA,ALGORITMY2020-Hybrid_Parallel_Polycrystalline}.
We review the behavior of several existing models with different reaction
terms in the Allen-Cahn \cite{Allen-Cahn-orig} equation. Depending
on the particular variant of the model, limitations exist in terms
of the applicability to physically realistic situations, in terms
of the possible mathematical and numerical analysis, or both. We see
the reaction term as means for controlling the spatial distribution
of latent heat release during solidification. Its form can be adjusted
while the asymptotic behavior of the model remains valid \cite{Caginalp-Stefan_HeleShaw_PF,Benes-Asymptotics}.
We take advantage of this flexibility and propose a new variant of
the reaction term. The resulting model is compatible with the numerical
analysis performed in our paper \cite{arXiv-PhaseField-FVM-Convergence}
for the finite volume method (as long as anisotropy is not considered)
and it also achieves a very good quantitative agreement with experiments
in the numerically difficult case of rapid solidification. In a number
of simulations, we compare the behavior of the discussed models.

\section{Analysis of the existing models}

To support intuitive understanding and in agreement with the works
\cite{Caginalp-Stefan_HeleShaw_PF,Caginalp-Convergence-PF_SIF} and
\cite{Benes-Math_comp_aspects_solid}, we first introduce a sharp
interface description of pure substance solidification in Section
\ref{subsec:Stefan-problem}, namely the Stefan problem with surface
tension \cite{Gurtin-Stefan_problem}. In Section \ref{subsec:General-Form-of-PF},
we proceed with the diffuse interface formulation by means of the
phase field model and we outline the asymptotic correspondence between
the two approaches in Section \ref{subsec:Results-of-matched-asymptotics}.
These results together with the discussion of the properties of some
known variants of the reaction term in Section \ref{subsec:Variants-of-f}
motivate us to modify the model further in Section \ref{sec:Designing-Alternative-Reaction-Terms}.

For the sake of simplicity, we proceed from an isotropic setting despite
the fact that our simulations mainly use anisotropy in agreement with
physical reality. Note however that the aniso\-tropic case introduced
later in Section \ref{sec:Anisotropic-Version} can also be treated
by a similar procedure \cite{dissertation,Benes-Diffuse-Interface}.

\subsection{\label{subsec:Stefan-problem}Stefan problem with surface tension}

Consider a domain $\Omega\subset\R^{3}$ and the time interval $\J=\left(0,T\right)$.
At each time $t$, $\Omega$ is divided into the solid subdomain $\Omega_{s}\left(t\right)$
and the liquid subdomain $\Omega_{l}\left(t\right)=\Omega\backslash\Omega_{s}\left(t\right)$
by the phase interface $\Gamma\left(t\right)$. Further on, let us
introduce the notation in Table \ref{tab:Physical-quantities}.
\begin{table}
\caption{\label{tab:Physical-quantities}Physical quantities in the Stefan
problem with surface tension}

\centering{}%
\begin{tabular}{ccl}
\toprule 
Quantity & SI Units & Description\tabularnewline
\midrule
$u$ & $\text{K}$ & temperature\tabularnewline
$\rho$ & $\text{kg}\cdot\text{m}^{-3}$ & density\tabularnewline
$c$ & $\text{J}\cdot\text{kg}^{-1}\text{K}^{-1}$ & specific heat capacity\tabularnewline
$\lambda$ & $\text{W}\cdot\text{m}^{-1}\cdot\text{K}^{-1}$ & heat conductivity\tabularnewline
$L$ & $\text{J}\cdot\text{m}^{-3}$ & latent heat of fusion per unit vol.\tabularnewline
$u^{*}$ & $\text{K}$ & melting point\tabularnewline
$\sigma$ & $\text{J}\cdot\text{m}^{-2}$ & surface tension\tabularnewline
$\Delta s$ & $\text{J}\cdot\text{m}^{-3}\cdot\text{K}$ & entropy difference per unit volume\tabularnewline
$\mu$ & $\text{m}\cdot\text{s}^{-1}\cdot\text{K}^{-1}$ & interface mobility (see \cite{Wheeler-PF-Binary_Alloys,Wheeler-PF-Dendrites})\tabularnewline
$\alpha$ & $\text{m}^{-1}\cdot\text{s}$ & coef. of attachment kinetics $\alpha=\frac{\Delta s}{\mu\sigma}$\tabularnewline
$\beta$ & $\text{m}^{-1}\cdot\text{K}^{-1}$ & $\beta=\frac{\Delta s}{\sigma}$\tabularnewline
\bottomrule
\end{tabular}
\end{table}

The system of governing equations reads
\begin{align}
\rho c\frac{\partial u}{\partial t} & =\nabla\left(\lambda\nabla u\right) & \text{in }\J\times\Omega_{s}\left(t\right)\label{eq:Stefan-heat-equation}\\
 &  & \text{and }\J\times\Omega_{l}\left(t\right),\nonumber \\
\left.b_{c}\left(u\right)\right|_{\partial\Omega} & =0 & \text{on }\J\times\partial\Omega,\label{eq:Stefan-boundary condition}\\
\left.u\right|_{t=0} & =u^{*}-\Delta u_{\text{ini}} & \text{in }\Omega,\label{eq:Stefan-initial-condition}\\
\left.\lambda\frac{\partial u}{\partial n_{\Gamma}}\right|_{s}-\left.\lambda\frac{\partial u}{\partial n_{\Gamma}}\right|_{l} & =Lv_{\Gamma} & \text{on }\Gamma\left(t\right),\label{eq:Stefan-condition}\\
\beta\left(u-u^{*}\right) & =-\kappa_{\Gamma}-\alpha v_{\Gamma} & \text{on }\Gamma\left(t\right),\label{eq:Stefan-Gibbs-Thompson-relation}\\
\Omega_{s}\left(0\right) & =\Omega_{s,\text{ini}}.\label{eq:Stefan-initial-solid-subdomain}
\end{align}
The problem consists of the heat equation (\ref{eq:Stefan-heat-equation})
valid within both subdomains, the Stefan condition (\ref{eq:Stefan-condition})
expressing the discontinuity of the heat flux across the interface
$\Gamma\left(t\right)$ and the Gibbs-Thomson relation (\ref{eq:Stefan-Gibbs-Thompson-relation})
describing the normal velocity of the interface $v_{\Gamma}$ as a
function of both the supercooling and the interface mean curvature
$\kappa_{\Gamma}$. The conditions (\ref{eq:Stefan-initial-condition})
and (\ref{eq:Stefan-initial-solid-subdomain}) prescribe the uniform
initial supercooling $\Delta u_{\text{ini}}>0$ and the spatial distribution
of the solid and liquid phase. The boundary condition (\ref{eq:Stefan-boundary condition})
for temperature can be of Dirichlet type, prescribing the temperature
$u_{\partial\Omega}$ on $\partial\Omega$ by
\[
b_{c}\left(u\right)=u-u_{\partial\Omega},
\]
or of Neumann type, specifying the heat flux $\vec{g}$ through $\partial\Omega$
by
\[
b_{c}\left(u\right)=\left(\lambda\nabla u-\vec{g}\right)\cdot\vec{n}.
\]

\subsection{Dimensionless formulation}

It is convenient to work with the phase field model in dimensionless
form, which can be achieved by defining the quantities (recognized
by tilde over the respective symbols) according to Table \ref{tab:Dimensionless-quantities}.
These relations imply that the initial supercooling $\Delta u_{\text{ini}}$
scales the value of $\tilde{\beta}$, whereas the dimensionless melting
point is always $\tilde{u}^{*}=1$. Similarly, the dimensionless initial
temperature satisfies $\left.\tilde{u}\right|_{t=0}=0$ independently
of $\Delta u_{\text{ini}}$. Furthermore, we also obtain the dimensionless
normal velocity of the interface
\[
\tilde{v}_{\Gamma}\left(\tilde{t}\right)=\frac{t_{0}}{L_{0}}v\left(t\right)
\]
and the dimensionless interface mean curvature
\[
\tilde{\kappa}=L_{0}\kappa.
\]
The Gibbs-Thomson relation (\ref{eq:Stefan-Gibbs-Thompson-relation})
transformed into dimensionless quantities therefore remains in the
same form
\begin{equation}
\tilde{\beta}\left(\tilde{u}-\tilde{u}^{*}\right)=-\tilde{\kappa}_{\Gamma}-\tilde{\alpha}\tilde{v}_{\Gamma}.\label{eq:Gibbs-Thomson-dimensionless}
\end{equation}
From this point on, we will only deal with dimensionless quantities
unless stated otherwise. For better readability, the tildes over the
respective symbols will be omitted.
\begin{table}
\caption{\label{tab:Dimensionless-quantities}Dimensionless quantities}

\centering{}%
\begin{tabular}{cll}
\toprule 
Dim-less qty & Definition & Description\tabularnewline
\midrule
$\tilde{u}$ & $\left(u-u^{*}\right)/\Delta u_{\text{ini}}+1$ & temperature\tabularnewline
$\tilde{L}$ & $L/\left(\rho c\Delta u_{\text{ini}}\right)$ & latent heat\tabularnewline
$L_{0}$ & user-defined & length scale\tabularnewline
$t_{0}$ & $\left(\rho c/\lambda\right)L_{0}^{2}$ & time scale\tabularnewline
$\tilde{\alpha}$ & $\left(\lambda/\left(\rho c\right)\right)\alpha$ & attachment kinetics coef.\tabularnewline
$\tilde{\beta}$ & $\beta L_{0}\Delta u_{\text{ini}}$ & \tabularnewline
$\tilde{\vec{x}}$ & $\vec{x}/L_{0}$ & dimensionless position\tabularnewline
$\tilde{t}$ & $t/t_{0}$ & dimensionless time\tabularnewline
\bottomrule
\end{tabular}
\end{table}

\subsection{\label{subsec:General-Form-of-PF}General form of the phase field
equations}

Let $p:\J\times\Omega\mapsto\left[0,1\right]$ be the phase field
equal to $0$ in the liquid phase and $1$ in the solid phase, with
a smooth transition in between, implicitly representing a diffuse
phase interface. Following \cite{Benes-Math_comp_aspects_solid,Caginalp-Stefan_HeleShaw_PF},
the system of governing equations in dimensionless form reads 
\begin{align}
\frac{\partial u}{\partial t} & =\Delta u+L\frac{\partial p}{\partial t} & \text{in }\J\times\Omega,\label{eq:heat-transfer-iso}\\
\alpha\xi^{2}\frac{\partial p}{\partial t} & =\xi^{2}\Delta p+f\left(u,p,\nabla p;\xi\right) & \text{in }\J\times\Omega,\label{eq:Allen-Cahn-iso}\\
\left.u\right|_{t=0} & =0,\;\left.p\right|_{t=0}=p_{\text{ini}} & \text{in }\Omega,\label{eq:initial-condition-u_p-iso}
\end{align}
coupled with either Dirichlet or homogeneous Neumann boundary conditions
chosen independently for $u$ and $p$, i.e.
\begin{equation}
\left.u\right|_{\partial\Omega}=u_{\partial\Omega}\text{ or }\nabla u\cdot\vec{n}=0\text{ on }\J\times\partial\Omega\label{eq:temperature-BC-iso}
\end{equation}
and
\begin{equation}
\left.p\right|_{\partial\Omega}=p_{\partial\Omega}\text{ or }\nabla p\cdot\vec{n}=0\text{ on }\J\times\partial\Omega.\label{eq:phase-field-BC-iso}
\end{equation}
Equation (\ref{eq:heat-transfer-iso}) is the heat equation with latent
heat release term, (\ref{eq:Allen-Cahn-iso}) is the Allen-Cahn equation
\cite{Allen-Cahn-orig}. The particular form of the reaction term
$f\left(u,p,\nabla p;\xi\right)$ in (\ref{eq:Allen-Cahn-iso}) distinguishes
between the individual variants of the model and will be discussed
later in Section \ref{subsec:Variants-of-f}. The role of the small
scalar parameter $\xi>0$ is shown below.

\subsection{\label{subsec:Results-of-matched-asymptotics}Results of the matched
asymptotic analysis}

Even though the phase field model (\ref{eq:heat-transfer-iso})--(\ref{eq:phase-field-BC-iso})
can be derived entirely from first principles \cite{Provatas_Elder-PF_Book},
our further steps benefit from some results of the mat\-ched asymptotic
analysis as $\xi\to0$. Its original purpose is to establish the correspondence
between the diffuse interface formulation in the phase field model
and the sharp interface formulation in the Stefan problem (\ref{eq:Stefan-heat-equation})--(\ref{eq:Stefan-initial-solid-subdomain}).
The procedure is explained in detail in \cite{Caginalp-Stefan_HeleShaw_PF}
and performed for the above model in \cite{Benes-Asymptotics}. Here
we summarize only those results that are relevant in our context.

Given the phase field $p$, the evolution of the sharp phase interface
$\Gamma$ can be determined implicitly by the relation
\begin{equation}
\Gamma\left(t\right)=\left\{ \left.\vec{x}\in\Omega\vphantom{\frac{1}{2}}\right|p\left(t,\vec{x}\right)=\frac{1}{2}\right\} .\label{eq:phase-interface}
\end{equation}

Let us consider the reaction term in the form (corresponding to the
expansion of $f$ into the powers of $\xi$ up to the first order)
\begin{equation}
f\left(u,p,\nabla p;\xi\right)=f_{0}\left(p\right)+\xi f_{1}\left(u,p,\nabla p;\xi\right)\label{eq:reaction-term-general-form}
\end{equation}
where 
\[
f_{0}\left(p\right)=2p\left(1-p\right)\left(p-\frac{1}{2}\right)
\]
represents the derivative of the double-well potential \cite{Caginalp-Stefan_HeleShaw_PF}

\[
\omega_{0}\left(p\right)=\frac{1}{2}\left(\left(p-\frac{1}{2}\right)^{2}-\frac{1}{4}\right)^{2}.
\]
The form of $\omega_{0}\left(p\right)$ is justified by the assumptions
taken in Landau theory \cite{Simple&ComplexLiq}. For $\xi$ sufficiently
small, it ensures that $f$ as a function of $p$ has roots in the
vicinity of $0$, $1$, and $\frac{1}{2}$, and that $p$ close to
zero is attracted to zero, whereas $p$ close to one is attracted
to one, giving rise to a thin interface layer in between.

\begin{figure}
\begin{centering}
\includegraphics[width=0.9\figwidth]{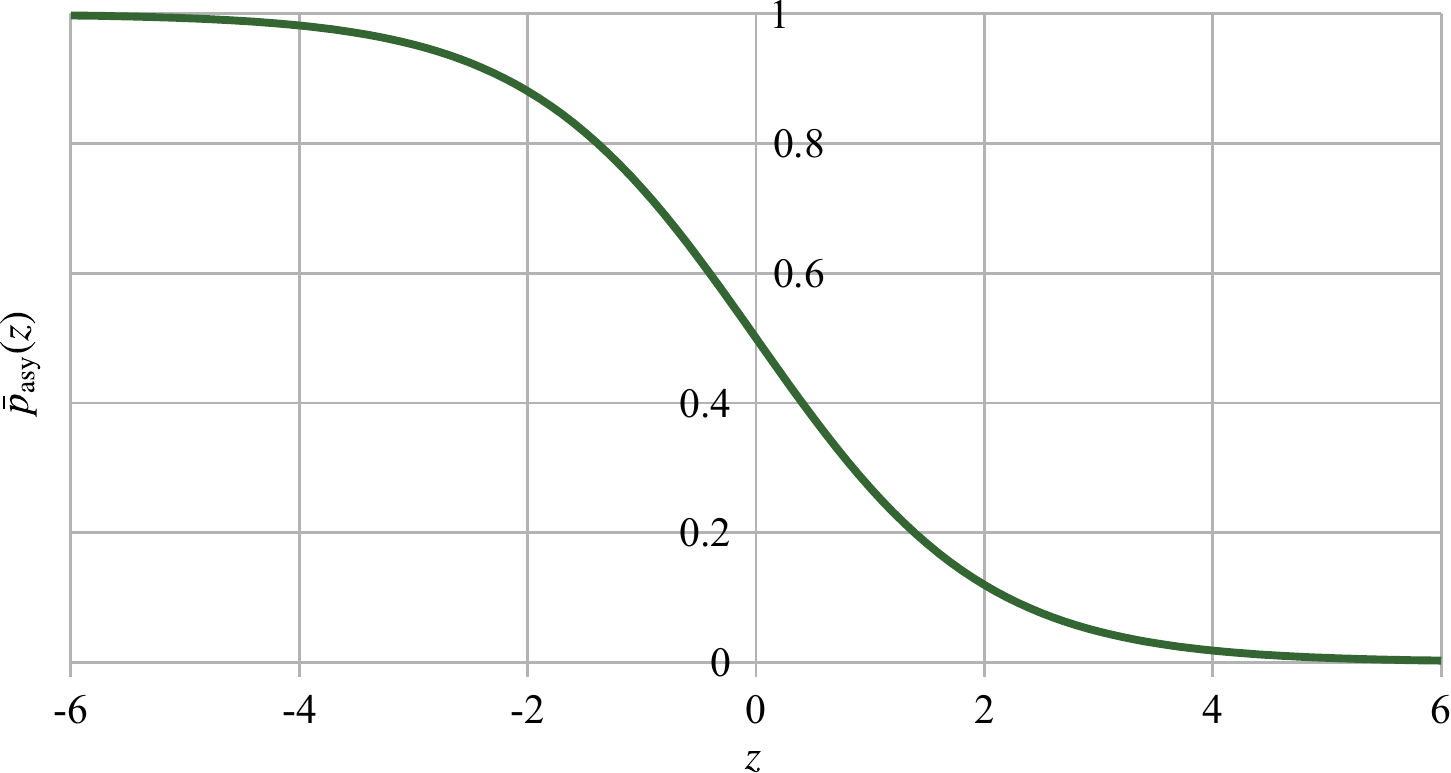}
\par\end{centering}
\caption{\label{fig:p_bar_profile}The asymptotic profile of the phase field
(the value of $\bar{p}_{\text{asy}}\left(z\right)$) across the diffuse
interface given by (\ref{eq:p_bar_profile}). A unit length in the
$z$ coordinate corresponds to the length $\xi$ in the original coordinates.}

\end{figure}

The inner asymptotic expansion indicates the profile of the phase
field $p$ across the interface $\Gamma\left(t\right)$ given by (\ref{eq:phase-interface}),
matching the values obtained from the outer expansion. In simple terms,
for $\vec{x}_{0}\in\Gamma\left(t\right)$, we introduce a coordinate
$z$ and a function $\bar{p}$ such that for a given time $t\in\J$,
we have
\[
\bar{p}\left(z\right)=p\left(t,\vec{x}_{0}+\xi z\vec{n}_{\Gamma}\right),
\]
$\vec{n}_{\Gamma}$ being the normal vector to $\Gamma$ at $\vec{x}_{0}$
pointing in the direction outside of the solid subdomain. Then the
function $\bar{p}$ satisfies
\begin{equation}
\bar{p}\left(z\right)=\bar{p}_{\text{asy}}\left(z\right)+o\left(\xi\right)\text{\;where\;}\bar{p}_{\text{asy}}\left(z\right)=\frac{1}{2}\left[1-\tanh\left(\frac{z}{2}\right)\right].\label{eq:p_bar_profile}
\end{equation}
The graph of $\bar{p}_{\text{asy}}$ is depicted in Figure \ref{fig:p_bar_profile}.
This graph indicates that the diffuse interface (where $\bar{p}$
is substantially different from both zero and one) effectively spans
a range $z\in\left(-5,5\right)$, yielding a thickness of approximately
$10\xi$ in the original coordinates. This is important for choosing
the mesh resolution in numerical simulations. Our simulations (see
Section \ref{sec:Computational-Studies}) confirm that it is usually
enough to have mesh element size equal to $\xi$, which results in
roughly $10$ mesh elements available to approximate $p$ across the
interface.

The other result important from the theoretical perspective is the
asymptotic recovery of the Gibbs-Thomson relation as $\xi\to0$. Depending
on the particular choice of $f_{1}$, this relation has the form \cite{Benes-Asymptotics}
\begin{equation}
\frac{\mathcal{I}_{1}}{\mathcal{I}_{2}}=-\kappa_{\Gamma}-\alpha v_{\Gamma}\label{eq:GT-relation-asymptotic}
\end{equation}
where
\[
\mathcal{I}_{1}=\int\limits _{-\infty}^{+\infty}f_{1}\left(\bar{u},\bar{p},\frac{\dd\bar{p}}{\dd z}\right)\frac{\dd\bar{p}}{\dd z}\dd z
\]
and the form of $f_{0}$ yields
\[
\mathcal{I}_{2}=\frac{1}{6}.
\]
In various forms of the term $f_{1}$, it is possible to tweak the
suitable constants so that (\ref{eq:GT-relation-asymptotic}) becomes
the Gibbs-Thomson relation of the Stefan problem (\ref{eq:Gibbs-Thomson-dimensionless}).

\subsection{\label{subsec:Variants-of-f}Some known variants of the reaction
term}

We review the formulation and properties of some existing reaction
terms. As they only differ in the choice of $f_{1}$ in (\ref{eq:reaction-term-general-form}),
all resulting models enjoy the same asymptotical properties discussed
in Section \ref{subsec:Results-of-matched-asymptotics}. However,
their practical applicability is limited. To overcome these limitations,
improved reaction term variants are designed later in Section \ref{sec:Designing-Alternative-Reaction-Terms}.

\subsubsection{The linear model}

Let us start with the simple version of the reaction term used in
\cite{Caginalp-Stefan_HeleShaw_PF}, namely
\begin{align}
f_{1}\left(u\right) & =b\beta\left(u^{*}-u\right),\label{eq:linear-model}\\
 & \Downarrow\nonumber \\
f\left(u,p;\xi\right) & =2p\left(1-p\right)\left(p-\frac{1}{2}\right)+\xi b\beta\left(u^{*}-u\right).\nonumber 
\end{align}
The recovery of the Gibbs-Thomson relation (\ref{eq:Gibbs-Thomson-dimensionless})
requires that $b=\mathcal{I}_{2}$. This model has two major drawbacks
that demonstrate themselves for practical simulations with $\xi$
fixed. First, it is immediately apparent that thanks to the form of
$f_{1}$, $f$ has nonzero values even for $p=0$ and $p=1$, so that
latent heat is released in the whole domain (not only at the interface)
until an equilibrium is achieved with values of $p$ shifted away
from $0$ and $1$. Second, the minimum of $f_{0}$ for $p\in\left[0,1\right]$
has the value $-\frac{\sqrt{3}}{18}$. As the initial dimensionless
supercooling $u^{*}-\left.u\right|_{t=0}$ is always equal to $1$,
the condition
\begin{equation}
\xi b\beta=\frac{1}{6}\xi\beta<\frac{\sqrt{3}}{18}\;\iff\;\xi\beta<\frac{\sqrt{3}}{3}\label{eq:condition-linear-model}
\end{equation}
is necessary for $f$ to maintain its three roots so that the interface
can develop its characteristic profile. For any realistic value of
$\beta$, this requires $\xi$ to be chosen very small and the mesh
resolution extremely fine (see Section \ref{subsec:Results-of-matched-asymptotics}
above), which is not possible even on contemporary computers, especially
for 3D simulations.

\subsubsection{The Kobayashi model}

In \cite{Kobayashi-PF-Dendritic}, the author proposes the reaction
term in the form
\begin{align}
f_{1}\left(u,p;\xi\right) & =2p\left(1-p\right)\frac{\alpha}{\pi\xi}\arctan\left(\gamma\left(u^{*}-u\right)\right),\label{eq:Kobayashi-model}\\
 & \Downarrow\nonumber \\
f\left(u,p;\xi\right) & =2p\left(1-p\right)\left(p-\frac{1}{2}+\frac{\alpha}{\pi}\arctan\left(\gamma\left(u^{*}-u\right)\right)\right).\nonumber 
\end{align}
For $\alpha\in\left(0,1\right)$, this model ensures that for any
$\gamma$, $f$ retains three roots: $p_{0}=0$, $p_{1}=1$, and the
third one $p_{2}\in\left(0,1\right)$. This means that latent heat
exchange is focused on the diffuse interface only. No latent heat
exchange occurs in pure solid and pure liquid. However, the use of
arctangent yields almost linear dependence of solidification speed
on supercooling (as given by (\ref{eq:Gibbs-Thomson-dimensionless}))
only for small values of $\gamma=\xi\beta$. With increasing values
of supercooling, this model remains well-posed, but is connection
with the Gibbs-Thomson relation is loosened.

\subsubsection{The gradient model (a.k.a. GradP model)}

In \cite{Benes-Math_comp_aspects_solid,Benes-Asymptotics}, the reaction
term is proposed in the form 
\begin{align}
f_{1}\left(u,p,\nabla p;\xi\right) & =\xi b\beta\left|\nabla p\right|\left(u^{*}-u\right),\label{eq:Gradient-model}\\
 & \Downarrow\nonumber \\
f\left(u,p,\nabla p;\xi\right) & =2p\left(1-p\right)\left(p-\frac{1}{2}\right)+\xi^{2}b\beta\left|\nabla p\right|\left(u^{*}-u\right)\nonumber 
\end{align}
and to recover the Gibbs-Thomson relation (\ref{eq:Gibbs-Thomson-dimensionless}),
it is calculated that $b=1$. This model has notable computational
advantages. It is obvious that $\left|\nabla p\right|=0$ when $p$
is constant, and therefore the term $f$ continues to have three roots
almost exactly at $0,1,\frac{1}{2}$ in the bulk solid and liquid.
The latent heat release is focused at the interface where the gradient
is nonzero. Several computational studies have been performed using
this model, both in 2D \cite{Benes-Comp_studies_Diff_Interface,Benes-Anisotropic_phase_field_model,Benes-Math_comp_aspects_solid,MMM2008}and
in 3D \cite{ENUMATH2011,ALGORITMY2016,ISPMA14-Pavel_Ales-ActaPhPoloA}.
However, for a very large supercooling, this model seems to be failing
as well (see Section \ref{subsec:Physically-Realistic-Simulations}).
The gradient term also introduces difficulties into the numerical
analysis of finite difference \cite{Benes-Comp_studies_Diff_Interface}
and finite volume \cite{ENUMATH2011} schemes.

\section{\label{sec:Designing-Alternative-Reaction-Terms}Designing alternative
reaction terms}

From the discussion in Section \ref{subsec:Results-of-matched-asymptotics},
it is obvious that the asymptotic limit $\xi\to0$ can never be properly
investigated by numerical simulations due to mesh resolution requirements,
especially in 3D. As long as the minimum value of $\xi$ is dictated
by the available computational resources, Section \ref{subsec:Variants-of-f}
highlights that the individual existing models have different properties
and limitations. In the following, we try to design a novel form of
the reaction term with two main objectives:
\begin{enumerate}
\item In order to broaden its practical utility, the model should be applicable
to simulations with very large values of supercooling, beyond the
capabilities of the GradP model.
\item The use of the computational algorithm based on the model should be
justified by numerical analysis. In our recent work \cite{arXiv-PhaseField-FVM-Convergence},
which was performed concurrently with the research presented here,
we prove convergence of the numerical solution to the unique solution
of the original problem with a rather generic form of the reaction
term $f$ which is \emph{local}, i.e. depending on $p$ and $u$ only.
The aim is therefore to avoid the dependence of $f$ on $\nabla p$.
\end{enumerate}

\subsection{\label{subsec:SigmaP1-P-Model}The $\Sigma$P1-P model}

The forms of the reaction term $f_{1}$ differ in the way how the
thermodynamic driving force (and the latent heat release) is distributed
or focused in the spatial domain $\Omega$. As the GradP model (\ref{eq:Gradient-model})
is known to work well in a relatively wide range of supercooling values
and other parameter settings, we first try to construct a local term
that would provide a similar distribution of latent heat release.

We recall (\ref{eq:p_bar_profile}), which relates the profile of
the phase field $\bar{p}$ across the diffuse interface and its asymptotic
limit $\bar{p}_{\text{asy}}$. As the gradient of the solution $p$
which for a given $t\in\J$ and $\vec{x}_{0}\in\Gamma\left(t\right)$
satisfies
\begin{equation}
\left|\nabla p\left(\vec{x}_{0}\right)\right|=\left|\frac{\partial p}{\partial\vec{n}_{\Gamma}}\left(\vec{x}_{0}\right)\right|=\left|\frac{1}{\xi}\frac{\dd\bar{p}}{\dd z}\left(0\right)\right|,\label{eq:grad-p-abs-val}
\end{equation}
a natural idea is to replace $\bar{p}$ by $\bar{p}_{\text{asy}}$
in (\ref{eq:grad-p-abs-val}). Directly from the matched asymptotic
analysis (see \cite{Benes-Asymptotics} for details) or by differentiation
of (\ref{eq:p_bar_profile}), it follows that $\bar{p}_{\text{asy}}$
satisfies the differential equation
\[
\frac{\dd\bar{p}_{\text{asy}}}{\dd z}=-2\bar{p}_{\text{asy}}\left(1-\bar{p}_{\text{asy}}\right).
\]
This leads us to substitute $\frac{1}{\xi}2p\left(1-p\right)$ (which
is non-negative for all $p\in\left[0,1\right]$) for $\left|\nabla p\right|$
in (\ref{eq:Gradient-model}) and arrive at the reaction term in the
form
\begin{align}
f_{1}\left(u,p\right) & =b\beta2p\left(1-p\right)\left(u^{*}-u\right),\label{eq:P1-P_model}\\
 & \Downarrow\nonumber \\
f\left(u,p;\xi\right) & =2p\left(1-p\right)\left(p-\frac{1}{2}+\xi b\beta\frac{1}{2}\left(u^{*}-u\right)\right).\nonumber 
\end{align}
In principle, an equivalent formula has already been mentioned in
\cite{Kobayashi-PF-Dendritic} as a precursor of the Kobayashi model
(\ref{eq:Kobayashi-model}).

Evaluating the asymptotic Gibbs-Thomson relation (\ref{eq:GT-relation-asymptotic}),
we first find that
\[
\mathcal{I}_{1}=b\beta\left(u^{*}-u\right)\int\limits _{0}^{1}p\left(1-p\right)\dd p=b\beta\mathcal{I}_{2}\left(u^{*}-u\right),
\]
which implies the recovery of the original Gibbs-Thomson relation
(\ref{eq:Gibbs-Thomson-dimensionless}) for $b=1$, as in the gradient
model (\ref{eq:Gradient-model}).

The reaction term $f$ given by (\ref{eq:P1-P_model}) has the roots
$p_{0}=0$, $p_{1}=1$ just like in the Kobayashi model (\ref{eq:Kobayashi-model}).
However, the existence of the third root $p_{2}\in\left(0,1\right)$
and hence the proper formation of the diffuse interface is bound by
the condition
\[
\xi\beta<1
\]
which is similar to (\ref{eq:condition-linear-model}) and only satisfied
for small values of supercooling. To relax this condition, we propose
a modified version
\begin{align}
f_{1}\left(u,p\right) & =b\beta\Sigma\left(p;\varepsilon_{0},\varepsilon_{1}\right)p\left(1-p\right)\left(u^{*}-u\right),\label{eq:SigmaP1-P_model}\\
 & \Downarrow\nonumber \\
f\left(u,p;\xi\right) & =2p\left(1-p\right)\left(p-\frac{1}{2}+\xi b\beta\frac{1}{2}\Sigma\left(p;\varepsilon_{0},\varepsilon_{1}\right)\left(u^{*}-u\right)\right),\nonumber 
\end{align}
where $\Sigma\left(p;\varepsilon_{0},\varepsilon_{1}\right)$ is a
differentiable sigmoid function (a limiter) in the form
\begin{equation}
\Sigma\left(p;\varepsilon_{0},\varepsilon_{1}\right)=\begin{cases}
0 & p\leq\varepsilon_{0},\\
1 & p\geq\varepsilon_{1},\\
\frac{3\left(p-\varepsilon_{0}\right)^{2}}{\left(\varepsilon_{1}-\varepsilon_{0}\right)^{2}}-\frac{2\left(p-\varepsilon_{0}\right)^{3}}{\left(\varepsilon_{1}-\varepsilon_{0}\right)^{3}} & p\in\left(\varepsilon_{0},\varepsilon_{1}\right).
\end{cases}\label{eq:Sigma-limiter}
\end{equation}
By multiplying the thermodynamic force term by $\Sigma$ which is
(close to) zero for $p\approx0$, the existence of the third root
$p_{2}\in\left(0,1\right)$ and thus the expected behavior of the
model is ensured independently of $\beta$, i.e. for virtually any
value of supercooling. The choice of $\varepsilon_{0},\varepsilon_{1}$
that works properly in the numerical algorithm will be discussed further
in Section \ref{subsec:Choice-of-Eps}.

\begin{figure}
\begin{centering}
\includegraphics[width=0.99\figwidth]{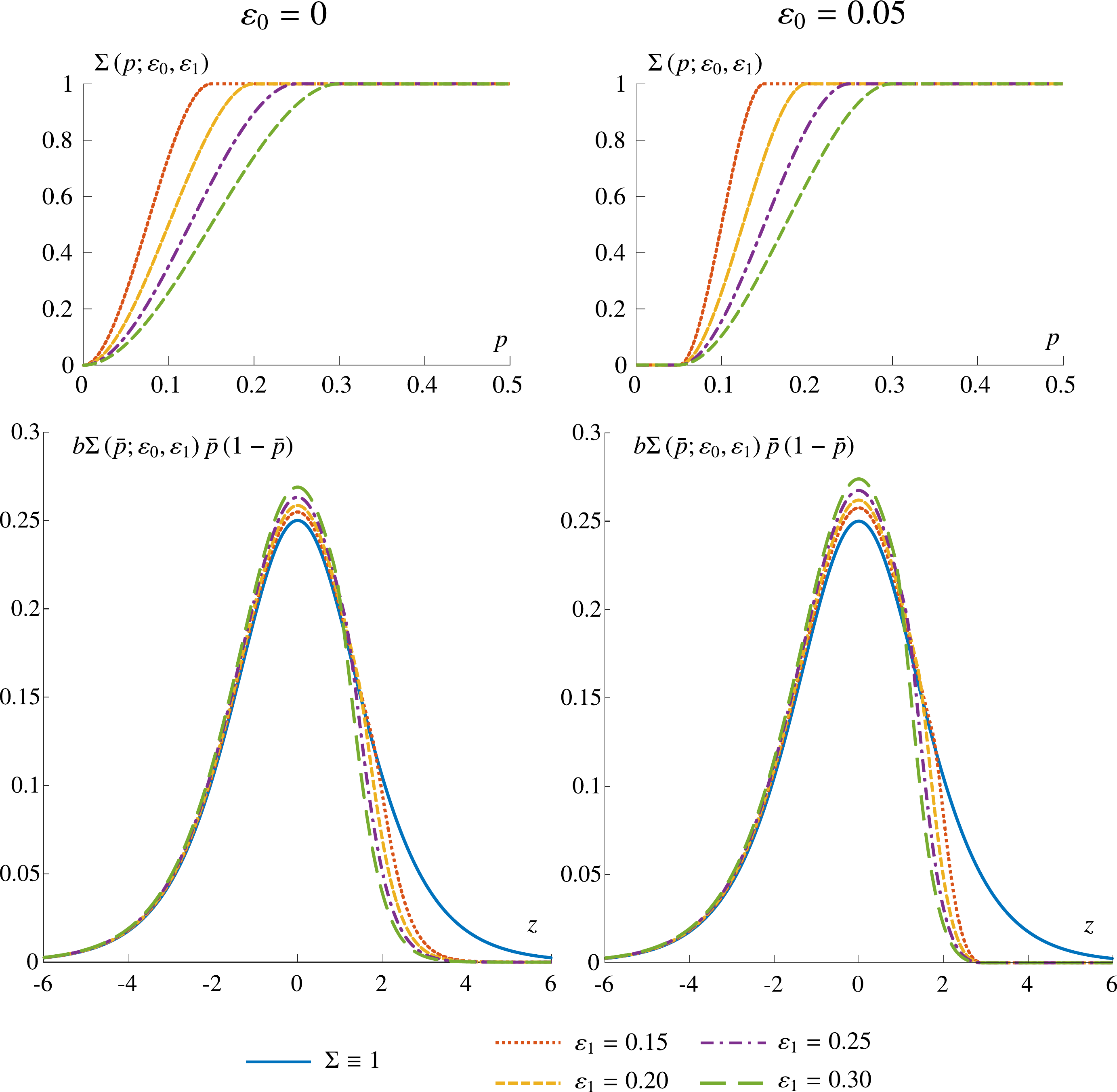}
\par\end{centering}
\caption{\label{fig:SigmaP1-P_profiles}Plots of $\Sigma\left(p;\varepsilon_{0},\varepsilon_{1}\right)$
for different values of $\varepsilon_{0},\varepsilon_{1}$ (top row)
and of the term $b\Sigma\left(\bar{p}\left(z\right);\varepsilon_{0},\varepsilon_{1}\right)\bar{p}\left(z\right)\left(1-\bar{p}\left(z\right)\right)$
which is proportional to the latent heat release rate across the asymptotic
profile of the diffuse interface (\ref{eq:p_bar_profile}) (bottom
row). The value of $b$ depends on $\varepsilon_{0},\varepsilon_{1}$
as given by (\ref{eq:b_COMPENSATION}).}
\end{figure}

In order to asymptotically satisfy (\ref{eq:Gibbs-Thomson-dimensionless}),
we calculate
\[
\mathcal{I}_{1}=b\beta\left(u^{*}-u\right)\int\limits _{0}^{1}\Sigma\left(p;\varepsilon_{0},\varepsilon_{1}\right)p\left(1-p\right)\dd p
\]
After plugging the result into (\ref{eq:GT-relation-asymptotic}),
we find that $b$ needs to be chosen as{\small{}
\begin{equation}
b=\frac{1}{6}\big/\left(\frac{\varepsilon_{0}^{3}}{15}+\frac{\varepsilon_{0}^{2}\varepsilon_{1}}{10}-\frac{3\varepsilon_{0}^{2}}{20}-\frac{\varepsilon_{0}\varepsilon_{1}}{5}+\frac{\varepsilon_{0}\varepsilon_{1}^{2}}{10}-\frac{3\varepsilon_{1}^{2}}{20}+\frac{\varepsilon_{1}^{3}}{15}+\frac{1}{6}\right).\label{eq:b_COMPENSATION}
\end{equation}
}The shape of the $\Sigma$ limiter (\ref{eq:Sigma-limiter}) and
the corresponding profiles of latent heat release across the diffuse
interface are plotted in Figure \ref{fig:SigmaP1-P_profiles}.

\subsection{The $\Sigma$GradP model}

The gradient model (\ref{eq:Gradient-model}) can also be modified
by incorporating $\Sigma$ and obtaining the reaction term in the
form{\small{}
\begin{align}
f_{1}\left(u,p,\nabla p;\xi\right) & =\xi b\beta\Sigma\left(p;\varepsilon_{0},\varepsilon_{1}\right)\left|\nabla p\right|\left(u^{*}-u\right),\label{eq:SigmaGradP-model}\\
 & \Downarrow\nonumber \\
f\left(u,p,\nabla p;\xi\right) & =2p\left(1-p\right)\left(p-\frac{1}{2}\right)+\xi^{2}b\beta\Sigma\left(p;\varepsilon_{0},\varepsilon_{1}\right)\left|\nabla p\right|\left(u^{*}-u\right).\nonumber 
\end{align}
}This modification may contribute to the stability of the phase interface
when large supercooling is used and the gradient model ceases to work
correctly. Some results calculated using the reaction term (\ref{eq:SigmaGradP-model})
are shown in Section \ref{subsec:Physically-Realistic-Simulations}.

\subsection{\label{subsec:Modeling-Irregular-Growth}Modeling irregular growth}

To obtain more realistic simulation results, random noise that modifies
the reaction term can be used \cite{Kobayashi-PF-Dendritic}. Loosely
inspired e.g. by \cite{KarmaRappel-Thermal_noise}, we propose to
use a time independent continuous noise field
\begin{equation}
\hat{u}:\Omega\to\left[-\frac{1}{2},\frac{1}{2}\right]\label{eq:u-noise-field}
\end{equation}
satisfying
\[
\int\limits _{\Omega}\hat{u}\left(\vec{x}\right)\dd x=0
\]
and introduce thermal fluctuations to the reaction term by replacing
$f\left(u,p,\nabla p;\xi\right)$ by $f\left(u+\delta\hat{u},p,\nabla p;\xi\right)$
in (\ref{eq:Allen-Cahn-iso}). The perturbation amplitude is controlled
by the parameter $\delta\geq0$.

\section{\label{sec:Anisotropic-Version}The anisotropic phase field model}

To incorporate anisotropic surface energies into the model, we follow
the approach of Finsler geometry as introduced by \cite{Bellettini_Paolini-Anis_motion_Finsler}
and later used in a number of our previous works, e.g. \cite{Benes-Anisotropic_phase_field_model,Benes-Comp_studies_Diff_Interface,Benes-Diffuse-Interface,ALGORITMY2009,ALGORITMY2016,ENUMATH2011,FVCA6-2011}.

The anisotropic phase field model arises by replacing equation (\ref{eq:Allen-Cahn-iso})
by
\begin{equation}
\alpha\xi^{2}\frac{\partial p}{\partial t}=\xi^{2}\nabla\cdot T^{0}\left(\nabla p\right)+f\left(u,p,\nabla p;\xi\right)\qquad\text{in }\J\times\Omega\label{eq:Allen-Cahn-anis}
\end{equation}
and the boundary condition (\ref{eq:phase-field-BC-iso}) by
\begin{equation}
T^{0}\left(\nabla p\right)\cdot\vec{n}=0\qquad\text{on }\J\times\partial\Omega.\label{eq:Neumann-BC-p-anis}
\end{equation}
The anisotropic operator $T^{0}$ (see \cite{Bellettini_Paolini-Anis_motion_Finsler,Benes-Anisotropic_phase_field_model,Benes-Comp_studies_Diff_Interface})
is derived from the convex dual Finsler metric $\phi^{0}\left(\vec{\eta}^{*}\right)$,
$\vec{\eta}^{*}\in\R^{3}$ as
\begin{equation}
T^{0}\left(\vec{\eta}^{*}\right)=\phi^{0}\left(\vec{\eta}^{*}\right)\phi_{\eta}^{0}\left(\vec{\eta}^{*}\right)\text{ where }\phi_{\eta}^{0}=\left(\partial_{\eta_{1}^{*}}\phi^{0},\partial_{\eta_{2}^{*}}\phi^{0},\partial_{\eta_{3}^{*}}\phi^{0}\right)^{\T}.\label{eq:T0-definition}
\end{equation}
The choice $\vec{\eta}^{*}=\nabla p$ in (\ref{eq:Allen-Cahn-anis})
and (\ref{eq:Neumann-BC-p-anis}) ensures that
\begin{equation}
\vec{n}=-\frac{\vec{\eta}^{*}}{\left|\vec{\eta}^{*}\right|}\label{eq:normal-vector}
\end{equation}
is the outer normal to $\Gamma$ for $\vec{x}\in\Gamma$.

If the function $\psi:\R^{3}\to\left(0,+\infty\right)$ represents
the anisotropic surface energy depending on the normal $\vec{n}$,
$\phi^{0}$ assumes the form \cite{Napolitano_Liu-3D_Crystal-melt_Wulff_shape,Gurtin-Thermomechanics_interfaces}
\begin{equation}
\phi^{0}\left(\vec{\eta}^{*}\right)=\left|\vec{\eta}^{*}\right|\psi\left(\vec{n}\right).\label{eq:phi0-in-the-model}
\end{equation}
For example, the formula for 4-fold anisotropy aligned with the specimen
coordinate system reads \cite{PunKay_Modeling_Anisotropic_Surface_Energies}
\begin{equation}
\psi\left(\vec{n}\right)=1+A_{1}\left[n_{1}^{4}+n_{2}^{4}+n_{3}^{4}-6\left(n_{1}^{2}n_{2}^{2}+n_{2}^{2}n_{3}^{2}+n_{3}^{2}n_{1}^{2}\right)\right]\label{eq:anisotropy-4fold}
\end{equation}
where the coefficient $A_{1}\ll1$ specifies the anisotropy strength.

\subsection{Reaction terms in the anisotropic model}

The reaction term $f$ in (\ref{eq:Allen-Cahn-anis}) can assume any
of the forms (\ref{eq:linear-model}), (\ref{eq:Kobayashi-model}),
(\ref{eq:Gradient-model}), (\ref{eq:SigmaP1-P_model}), (\ref{eq:SigmaGradP-model}).
In addition, the gradient model (\ref{eq:Gradient-model}) and the
$\Sigma$GradP model (\ref{eq:SigmaGradP-model}) can be further modified
by replacing $\left|\nabla p\right|$ by the anisotropic norm $\phi^{0}\left(\nabla p\right)$.
In practice, this modification slightly increases the solidification
rate where the outer normal to the solid-liquid interface points in
the preferred directions of dendrite growth. The resulting models
will be referred to as $\phi^{0}$GradP and $\Sigma\phi^{0}$GradP,
respectively. Noise can be added to (\ref{eq:Allen-Cahn-anis}) exactly
as described in Section \ref{subsec:Modeling-Irregular-Growth}.

\section{\label{sec:Computational-Studies}Computational studies}

The primary purpose of the computational studies shown below is to
understand the behavior of the $\Sigma$P1-P model (\ref{eq:SigmaP1-P_model})
in comparison to the original GradP model (\ref{eq:Gradient-model})
and its anisotropic variant (the $\phi^{0}$GradP model). The interface
thickness (and $\xi$) is chosen as small as possible for the computational
costs to remain feasible on our hardware. This means that the asymptotic
limit for $\xi\to0$ is not pursued. The phase field model with aniso\-tropy
given by equations (\ref{eq:heat-transfer-iso}), (\ref{eq:Allen-Cahn-anis}),
(\ref{eq:initial-condition-u_p-iso}), (\ref{eq:temperature-BC-iso}),
and (\ref{eq:phase-field-BC-iso}) is used.

All simulations were performed using our efficient hybrid OpenMP/MPI
parallel implementation \cite{ALGORITMY2016} of the numerical solver
based on multipoint flux approximation finite volume scheme on a uniform
rectangular mesh \cite{ENUMATH2011} and 4th order Runge-Kutta-Merson
integrator with adaptive stepping in time \cite{Christiansen-RK-Merson}.
Note that keeping the random noise field (\ref{eq:u-noise-field})
constant in time guarantees that the time step adjustment algorithm
is not affected.

\subsection{\label{subsec:Basic-simulations-setup}Initial \& boundary conditions,
parameter settings}

The domain $\Omega$ is a cube in the form $\Omega=\left(0,\ell\right)^{3}$
discretized by a uniform mesh of $N\times N\times N$ cells. The values
of $\ell$ and $N$ together with the anisotropy specification vary
and are reported separately for each of the studies. Zero Neumann
boundary conditions are chosen for both $u$ and $p$ which allows
to initiate the solidification at a spherical nucleation site $\Omega_{s}\left(0\right)$
located either in the center or in the corner of the domain $\Omega$.
In the latter case, $\Omega$ represents one octant of a larger domain
where centrally and axially symmetric solidification takes place.
The initial condition (\ref{eq:initial-condition-u_p-iso}) for the
phase field is given by
\[
p_{\text{ini}}\left(\vec{x}\right)=\begin{cases}
1 & \vec{x}\in\Omega_{s}\left(0\right),\\
0 & \vec{x}\in\Omega_{l}\left(0\right)=\Omega\backslash\Omega_{s}\left(0\right).
\end{cases}
\]
Except in Section \ref{subsec:Physically-Realistic-Simulations},
a reference set of parameters comparable to the choices in \cite{Benes-Math_comp_aspects_solid,Kobayashi-PF-Dendritic}
is specified as $\alpha=3$, $\beta=300$, $\xi=0.011$, $L=2$. Noise
is not used unless a positive value of $\delta$ is explicitly given.

\subsection{\label{subsec:Latent-heat-release--profiles}Latent heat release
profiles at the diffuse interface}

\begin{figure}
\centering{}\includegraphics[width=0.99\figwidth]{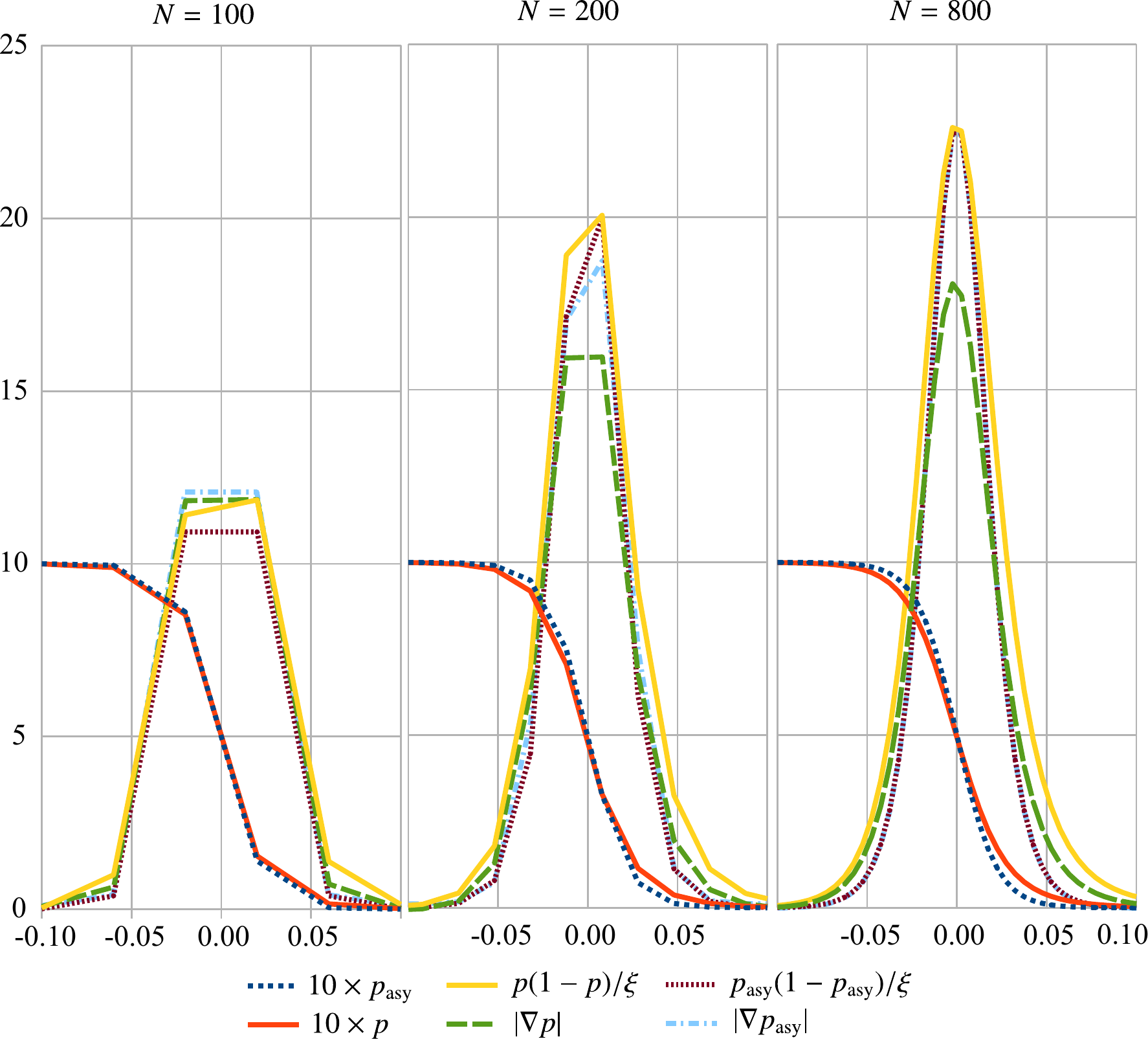}\caption{\label{fig:Interface-profiles}The computed profiles of $p$ across
the diffuse phase interface compared to the asymptotic profile $p_{\text{asy}}$
on meshes with different resolutions. The horizontal axis represents
the $\xi z$ coordinate, i.e. the signed distance from the phase interface.
Based on both $p$ and $p_{\text{asy}}$, the key quantities determining
the latent heat release profile across the interface are evaluated
for the GradP and the $\Sigma$P1-P models. The gradients are calculated
numerically by 4th order central differences also used in the numerical
scheme (see Section \ref{sec:Computational-Studies}), which explains
why $\left|\nabla p_{\text{asy}}\right|$ and $p_{\text{asy}}\left(1-p_{\text{asy}}\right)/\xi$
do not coincide.}
\end{figure}

In order to evaluate the difference between the GradP and the $\Sigma$P1-P
models in terms of latent heat release focusing, we extracted the
profiles of $p$ across the diffuse phase interface from numerical
simulations and compared them to the asymptotic profile $p_{\text{asy}}$.
The computations were performed in a domain of size $\ell=4$ with
mesh resolutions $N\in\left\{ 100,200,800\right\} $ and $\xi=0.011$.
We verified that the diffuse interface forms shortly after the start
of the simulation and the use of anisotropy, the location of readout
of the values of $p$ along the interface and the choice of the model
all have a negligible effect on its shape. Figure \ref{fig:Interface-profiles}
demonstrates the subtle difference between $p$ and $p_{\text{asy}}$
on different meshes. In addition, the quantities $\left|\nabla p\right|$
, $\frac{1}{\xi}2p\left(1-p\right)$ occurring in the GradP and the
$\Sigma$P1-P models, respectively, are plotted. 

\subsection{\label{subsec:Choice-of-Eps}Effect of $\varepsilon_{0},\varepsilon_{1}$
in the $\Sigma$P1-P model}

\begin{figure}
\begin{centering}
\includegraphics[width=0.9\figwidth]{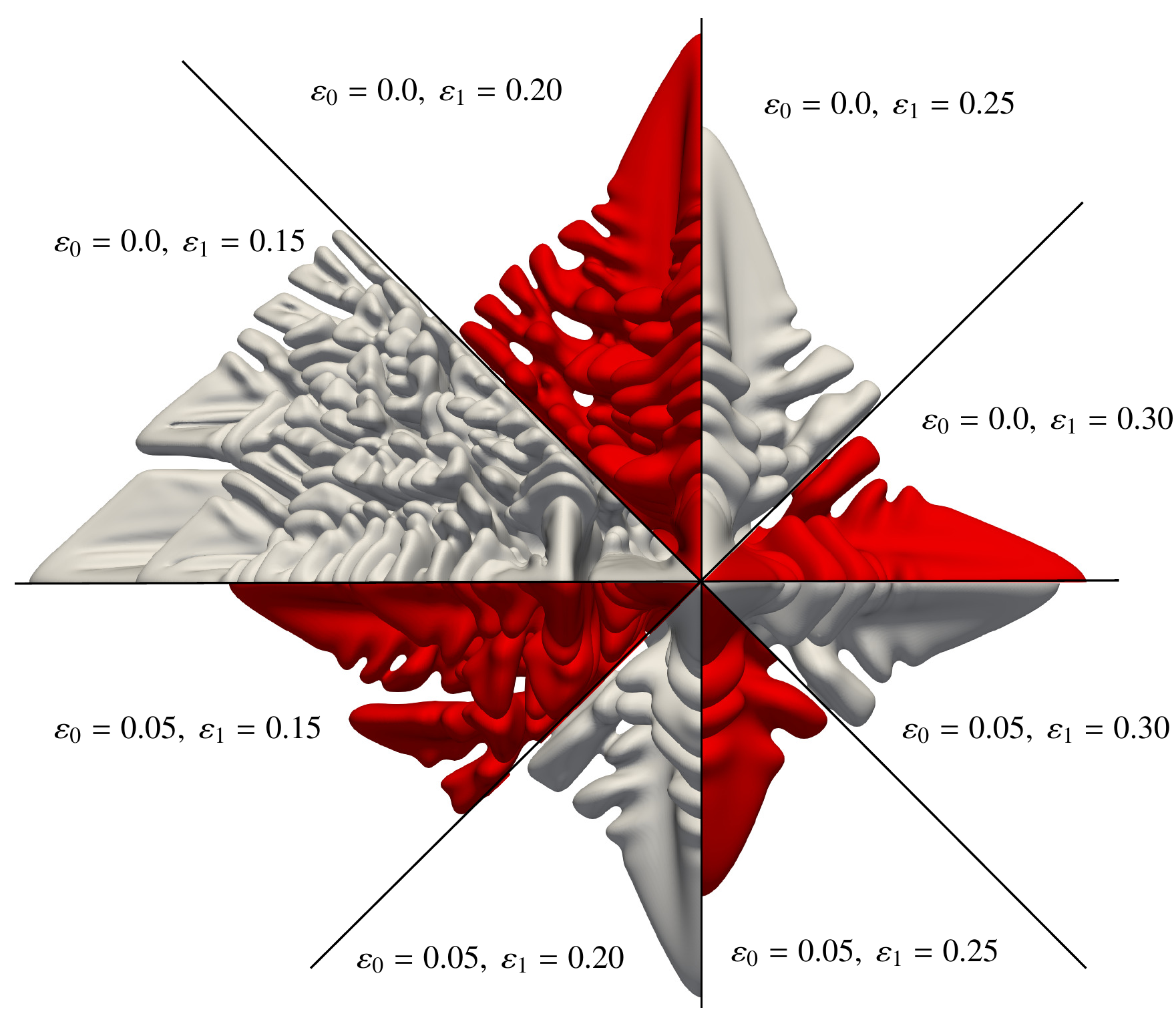}
\par\end{centering}
\caption{\label{fig:EPS_study}Influence of $\varepsilon_{0},\varepsilon_{1}$
on the crystal shape obtained by the $\Sigma$P1-P model with $b=1$.
Details are in Section \ref{subsec:Choice-of-Eps}.}

\end{figure}

\begin{figure}
\begin{centering}
\includegraphics[width=0.9\figwidth]{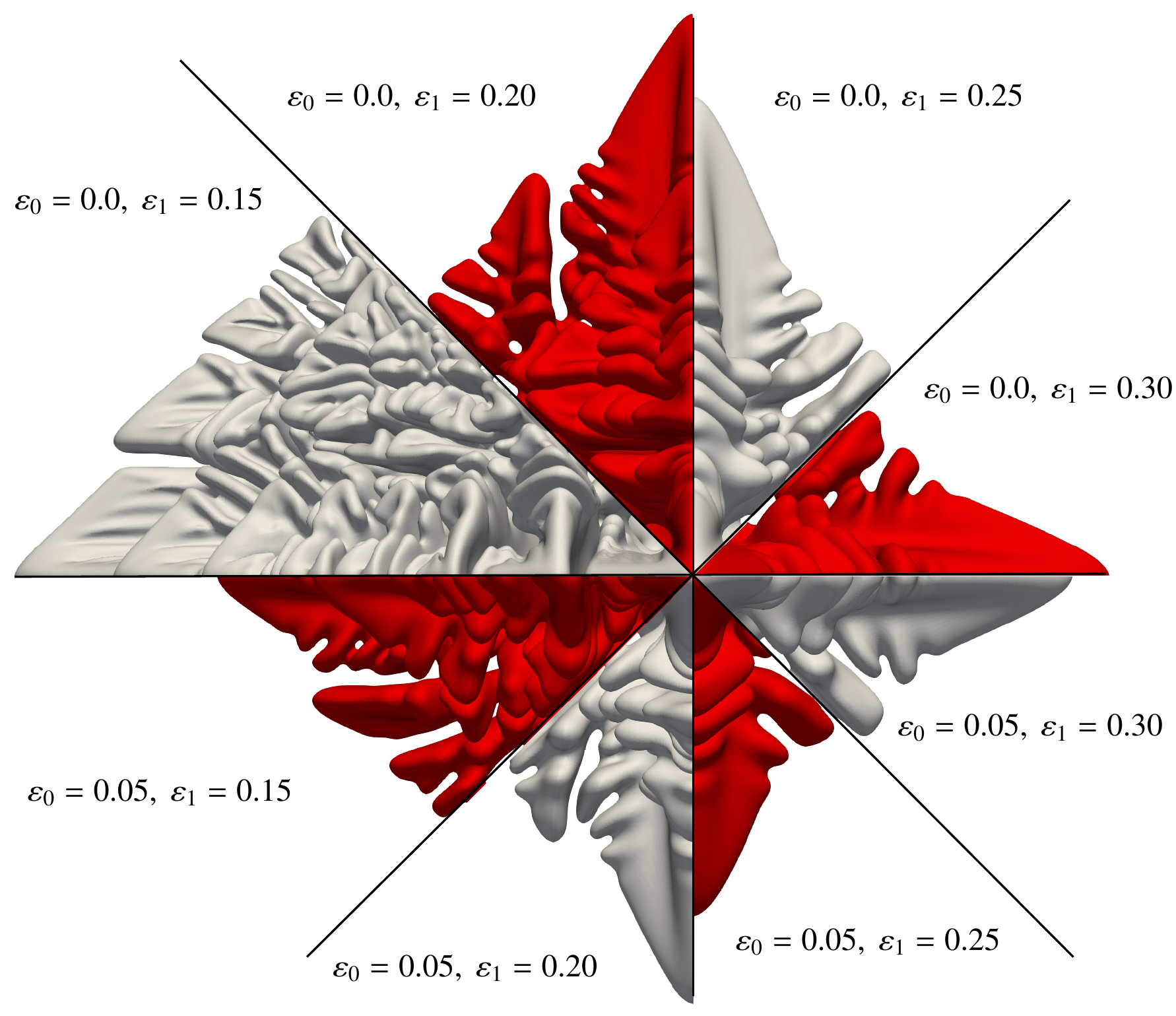}
\par\end{centering}
\caption{\label{fig:EPS_study-COMPENSATED}Influence of $\varepsilon_{0},\varepsilon_{1}$
on the crystal shape obtained by the $\Sigma$P1-P model with $b$
given by (\ref{eq:b_COMPENSATION}). Details are in Section \ref{subsec:Choice-of-Eps}.}
\end{figure}

To investigate the role of $\varepsilon_{0},\varepsilon_{1}$ in the
$\Sigma$P1-P model, simulations with
\[
\varepsilon_{0}\in\left\{ 0,0.05\right\} ,\;\varepsilon_{1}\in\left\{ 0.15,0.20,0.25,0.30\right\} 
\]
were performed, which corresponds to the solidification focusing profiles
shown in Figure \ref{fig:SigmaP1-P_profiles}. An even lower value
$\varepsilon_{1}=0.1$ was not enough for the diffuse interface to
form correctly. The results were obtained with $\ell=6$, $N=400$,
4-fold anisotropy given by (\ref{eq:anisotropy-4fold}) with $A_{1}=0.02$.
The nucleation site with radius $0.05$ was located in the corner
of $\Omega$ at $\vec{x}=\vec{0}$ and the snapshots at time $t=0.26$
were compared. As noise was not used, it is possible to demonstrate
only part of the whole crystal, the rest being given by symmetry.
By this approach, up to eight shapes can be compared in different
sectors of space viewed in the direction of the $z$ axis, as shown
in Figures \ref{fig:EPS_study} and \ref{fig:EPS_study-COMPENSATED}.

\begin{figure}
\begin{centering}
\includegraphics[width=0.8\figwidth]{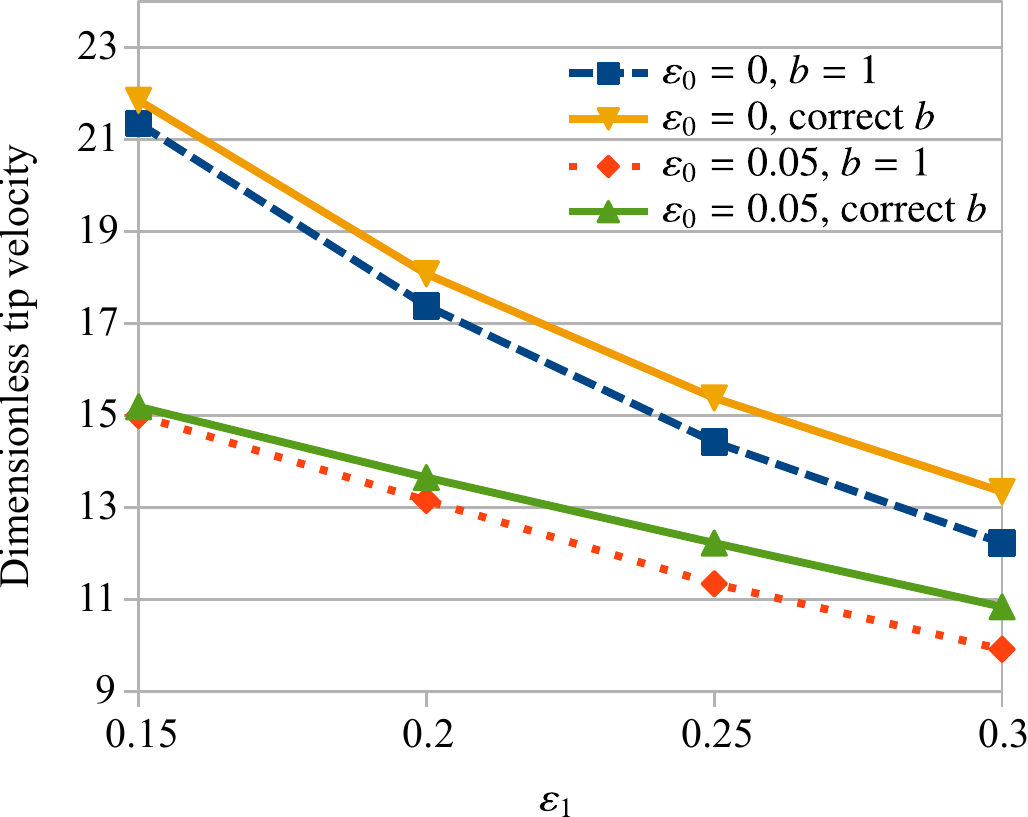}
\par\end{centering}
\caption{\label{fig:EPS_study-tip-velocity}Influence of $\varepsilon_{0},\varepsilon_{1}$
on the dimensionless dendrite tip velocity in the $\Sigma$P1-P model.}
\end{figure}

First, Figure \ref{fig:EPS_study} compares the crystal shapes with
$b$ in (\ref{eq:SigmaP1-P_model}) fixed to $1$ instead of being
given by (\ref{eq:b_COMPENSATION}). The model therefore deviates
from the Gibbs-Thomson condition (\ref{eq:Gibbs-Thomson-dimensionless})
and significant differences in results depending on the values of
$\varepsilon_{0},\varepsilon_{1}$ are obtained, as expected. In Figure
\ref{fig:EPS_study-COMPENSATED}, the value of $b$ is given by (\ref{eq:b_COMPENSATION}),
but nonetheless the results stay very similar to those in Figure \ref{fig:EPS_study}.
This observation testifies that the evolution of the complex dendritic
structure is very sensitive to changing the focusing of latent heat
release, despite the fact that the asymptotic behavior for $\xi\to0$
should be the same regardless of $\varepsilon_{0},\varepsilon_{1}$.
On the other hand, using the ``correct'' value of $b$ still slightly
compensates the drop in dendrite tip velocity with increasing $\varepsilon_{1}$,
as demonstrated in Figure \ref{fig:EPS_study-tip-velocity}.


\subsection{\label{subsec:Model-Comparison}Model comparison}

The next set of results illustrates the properties of the $\Sigma$P1-P
model both with $b=1$ and with $b$ given by (\ref{eq:b_COMPENSATION})
in comparison with the original GradP and $\phi^{0}$GradP models.

\subsubsection{\label{subsec:4-fold-Anisotropic-Simulations}4-fold anisotropic
simulations}

\begin{figure}
\begin{centering}
\includegraphics[width=0.9\figwidth]{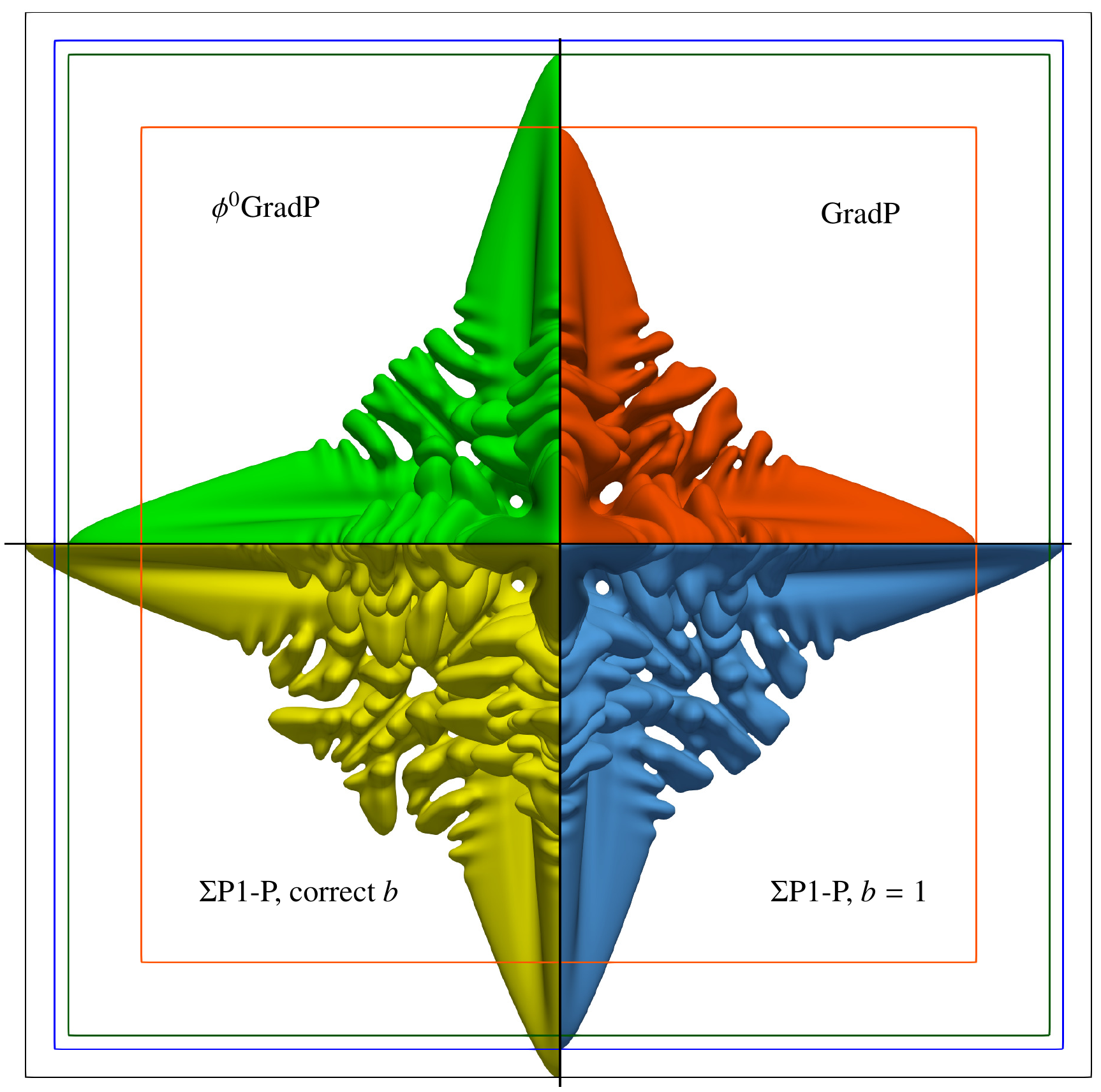}
\par\end{centering}
\caption{\label{fig:4-fold-model-comparison}Comparison of crystal shapes obtained
by the four reaction term models. The squares indicate the bounding
boxes for the individual crystals to assist the visual comparison
of average dendrite tip velocities. Details on simulation setup are
in Section \ref{subsec:4-fold-Anisotropic-Simulations}.}
\end{figure}

Figure \ref{fig:4-fold-model-comparison} demonstrates the solution
by all four models at time $t=0.32$. The problem settings are $\ell=6$,
$N=600$, 4-fold anisotropy given by (\ref{eq:anisotropy-4fold})
with $A_{1}=0.02$, and the nucleation site with radius $0.05$ located
in the corner of $\Omega$. The $\Sigma$P1-P model parameters read
$\varepsilon_{0}=0.05$, $\varepsilon_{1}=0.2$. For this setup, all
four models exhibit the same qualitative behavior with slightly different
dendrite tip velocities.

\begin{figure}
\begin{centering}
\includegraphics[width=0.9\figwidth]{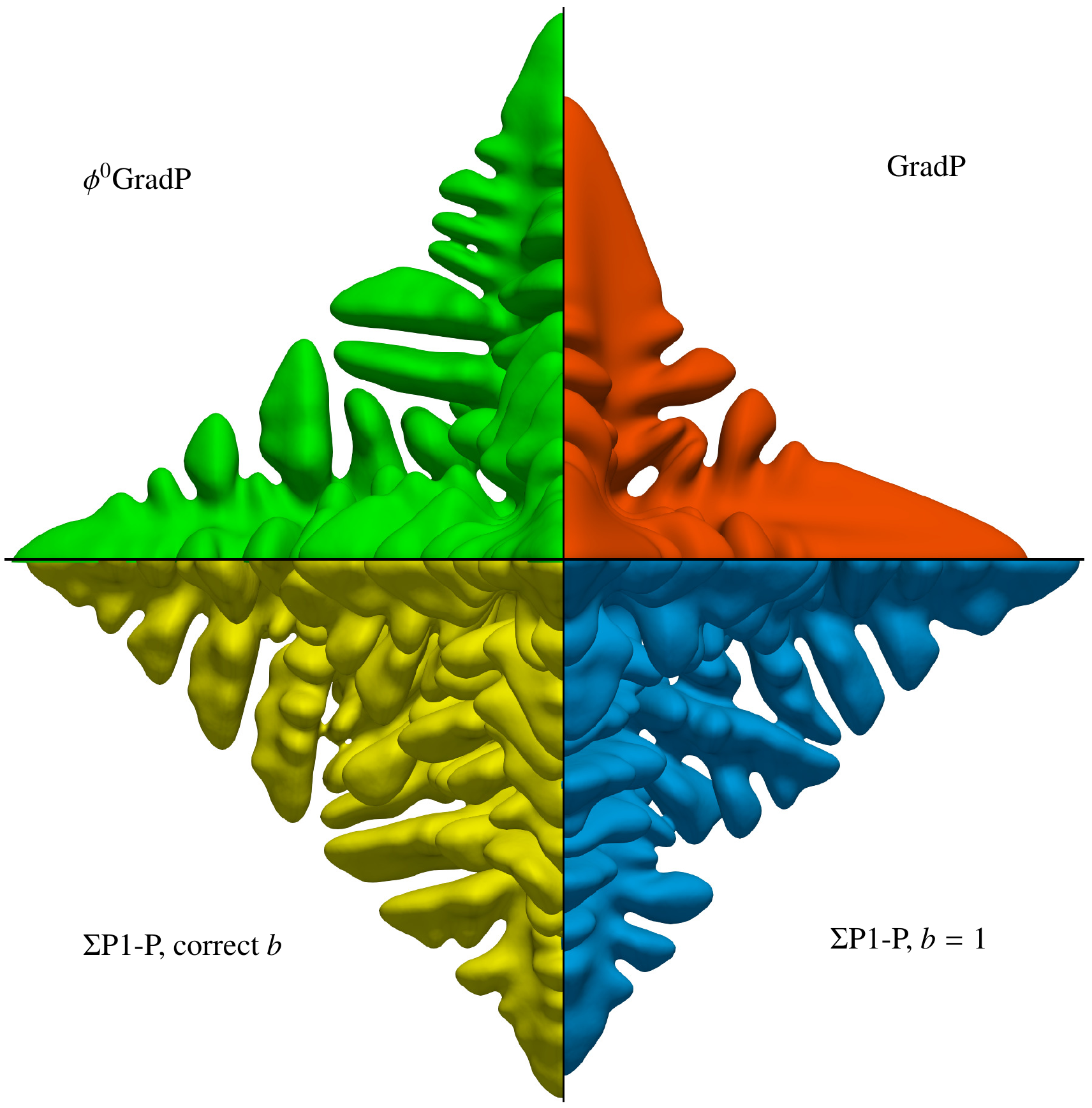}
\par\end{centering}
\caption{\label{fig:4-fold+NOISE-model-comparison}Comparison of crystal shapes
obtained by the four reaction term models with random fluctuations
of the temperature field. Details on simulation setup are in Section
\ref{subsec:4-fold-Anisotropic-Simulations}.}
\end{figure}

In Figure \ref{fig:4-fold+NOISE-model-comparison}, a similar comparison
is made at time $t=0.4$, this time with added noise by means of
(\ref{eq:u-noise-field}) with $\delta=0.05$. The other settings
are $\ell=8$, $N=600$, 4-fold anisotropy with $A_{1}=0.02$. Growth
symmetries are no longer valid with random thermal fluctuations present.
In order to see that, the nucleation site with radius $0.05$ is placed
in the center of $\Omega$. The resulting shapes are indeed asymmetric,
but only one octant is visualized for each model in Figure \ref{fig:4-fold+NOISE-model-comparison}.
Again, all models are qualitatively commensurate and the noise is
shown to induce side branching \cite{KarmaRappel-Thermal_noise}.

\begin{figure}
\begin{centering}
\includegraphics[width=0.9\figwidth]{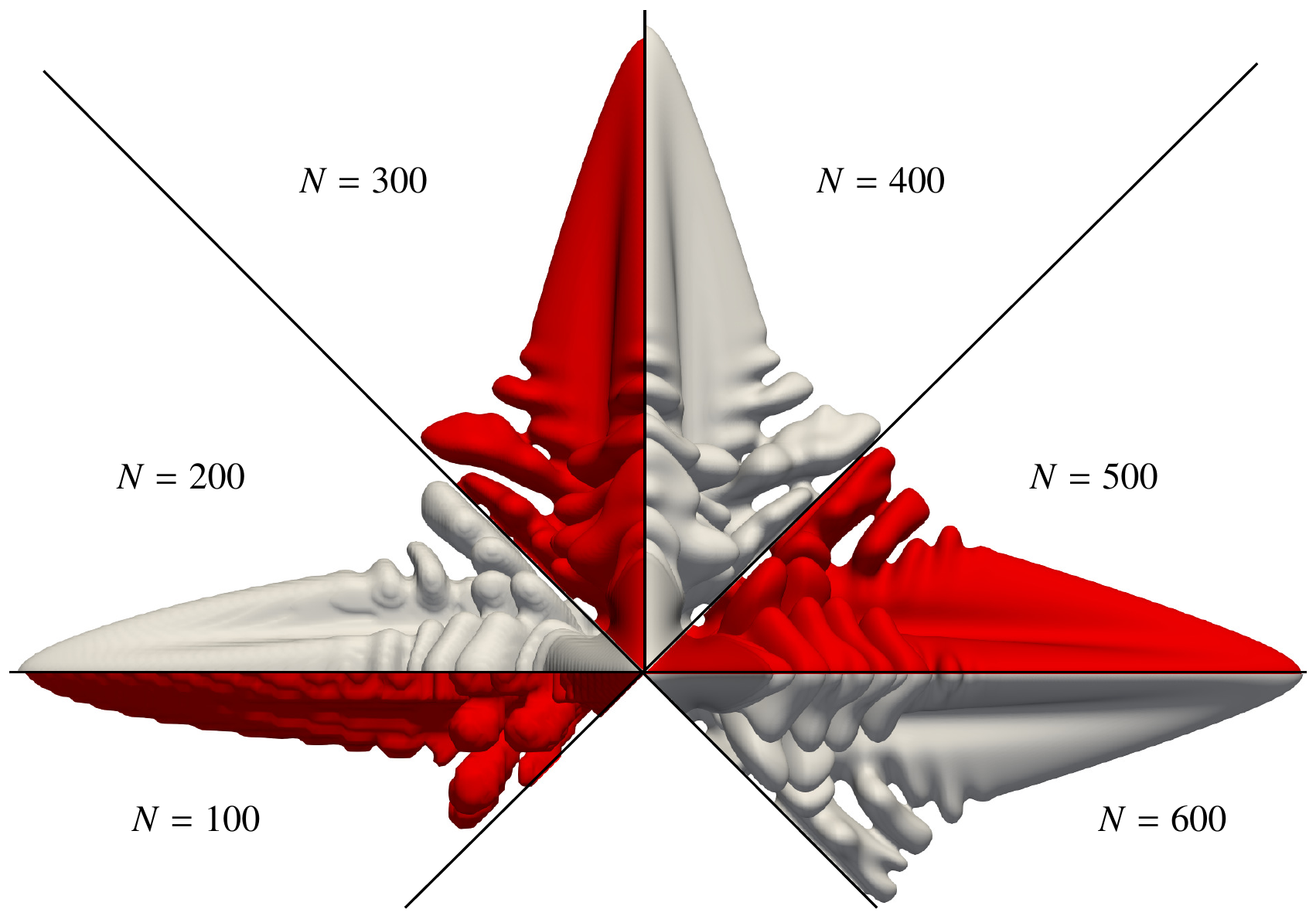}
\par\end{centering}
\caption{\label{fig:4-fold-Phi0GradP-mesh_dependence}Dependence of the crystal
shape on mesh resolution for the $\phi^{0}$GradP model.}
\end{figure}

\begin{figure}
\begin{centering}
\includegraphics[width=0.9\figwidth]{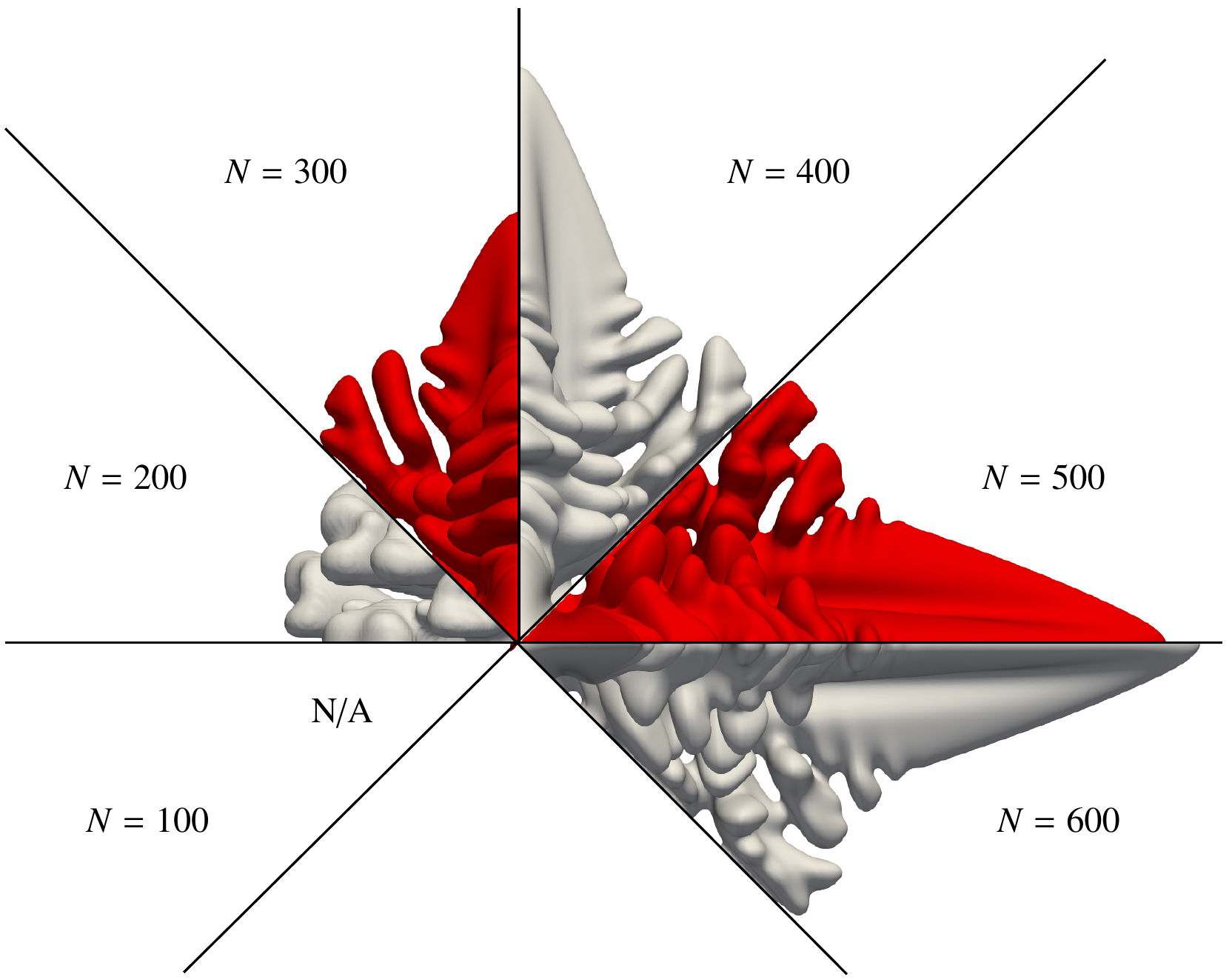}
\par\end{centering}
\caption{\label{fig:4-fold-SigmaP1-P-mesh_dependence}Dependence of the crystal
shape on mesh resolution for the $\Sigma$P1-P model with $b=1$.}
\end{figure}

Figures \ref{fig:4-fold-Phi0GradP-mesh_dependence} and \ref{fig:4-fold-SigmaP1-P-mesh_dependence}
demonstrate the dependence of the solution on the mesh resolution.
The parameters for the simulations are the same as in Figure \ref{fig:4-fold-model-comparison}
(the first result of this section) except for the changing value of
$N$. At lower resolutions, the $\phi^{0}$GradP model provides more
consistent results than the $\Sigma$P1-P model. The anisotropic growth
is correctly captured by both models from mesh resolution above $N=300$
and convergence of the solution in terms of the crystal shape can
be observed as $N$ approaches the finest resolution $N=600$.

\begin{figure}
\begin{centering}
\includegraphics[width=0.95\figwidth]{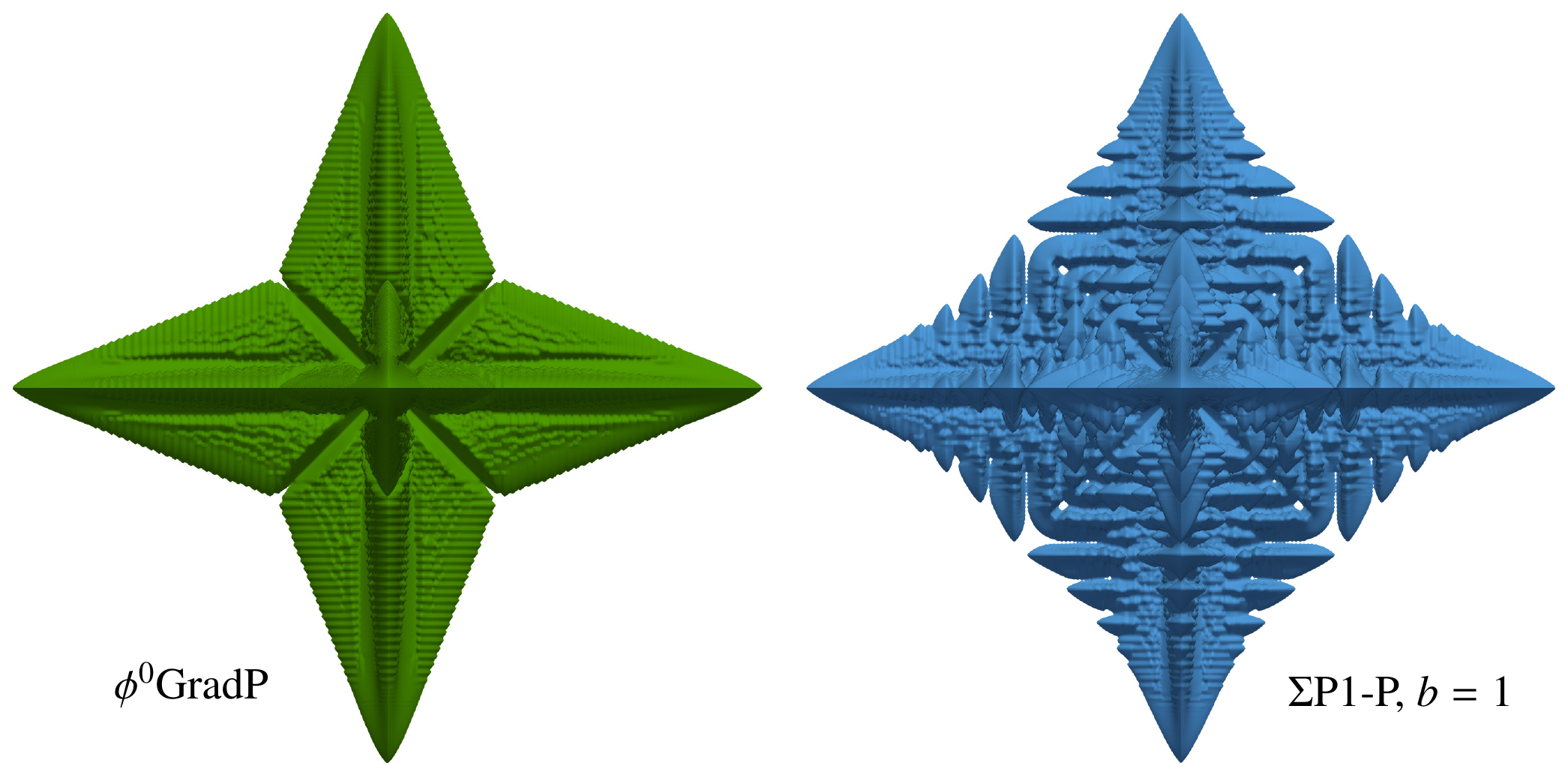}
\par\end{centering}
\caption{\label{fig:Crystalline}Comparison of crystal shapes produced by the
respective models for anisotropy strength beyond the convexity limit
for $\phi^{0}$. Both models produce faceted crystal surfaces. Details
on simulation setup are in Section \ref{subsec:4-fold-Anisotropic-Simulations}.}
\end{figure}

Finally, we demonstrate how the $\phi^{0}$GradP and $\Sigma$P1-P
models compare in a simulation with 4-fold anisotropy strength $A_{1}=0.2$,
which is a value large enough for $\phi^{0}$ to lose its convexity
property \cite{Bellettini_Paolini-Anis_motion_Finsler}. Such situation
is known to yield faceted crystal surfaces \cite{Gurtin-Thermomechanics_interfaces,Kobayashi-Anis_curvature_crystals}.
The problem settings are $\ell=4$, $N=200$, $t=0.13$ and the nucleation
site is located in the corner of $\Omega$. The $\Sigma$P1-P model
parameters are chosen as $\varepsilon_{0}=0.05$, $\varepsilon_{1}=0.2$,
$b=1$. In Figure \ref{fig:Crystalline}, it is visible that both
models produce sharp dendrite tips and facets.

\subsubsection{\label{subsec:Isotropic-Simulations}Isotropic Simulations}

\begin{figure}
\begin{centering}
\includegraphics[height=0.4\columnwidth]{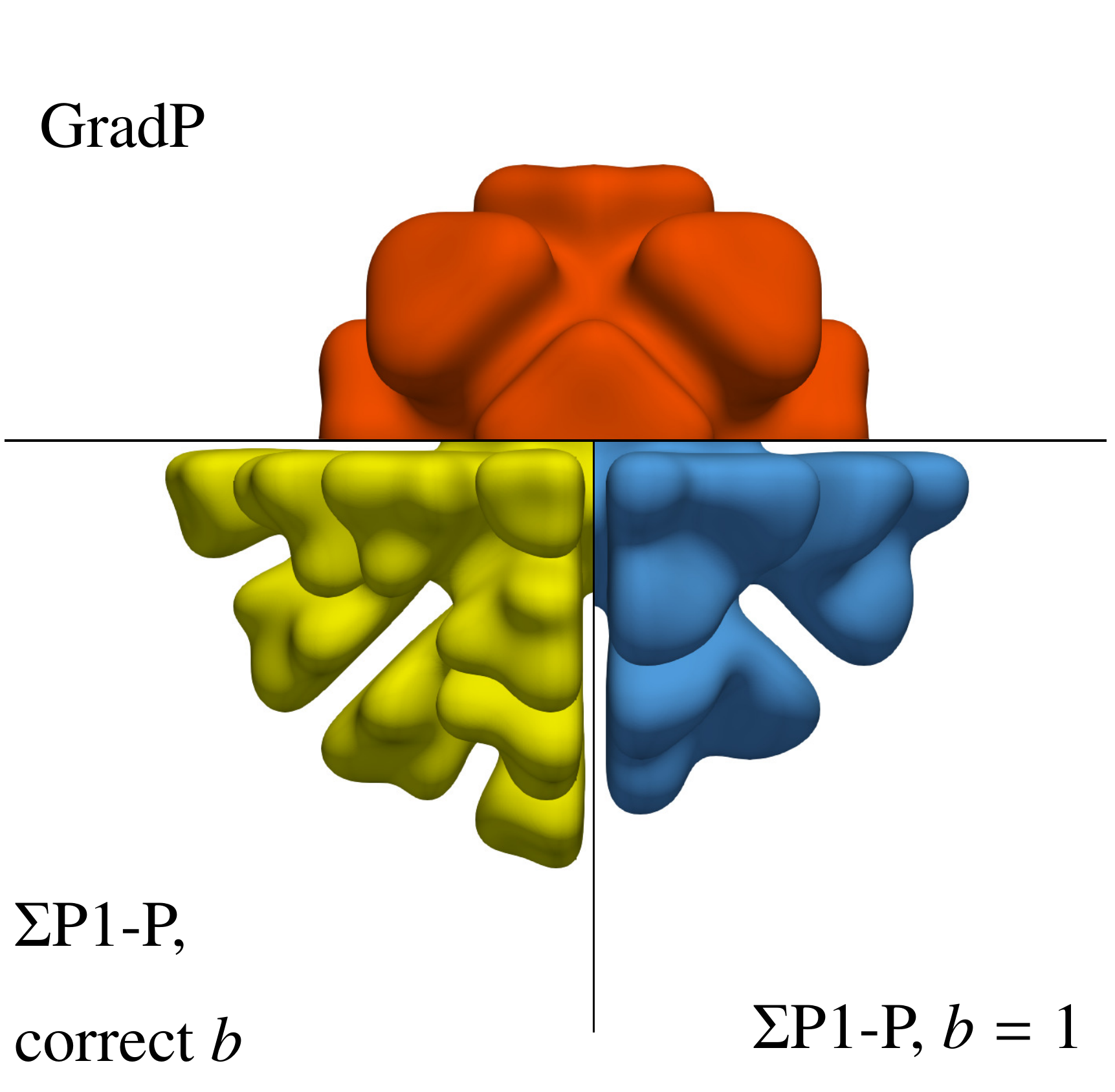}~\includegraphics[height=0.4\columnwidth]{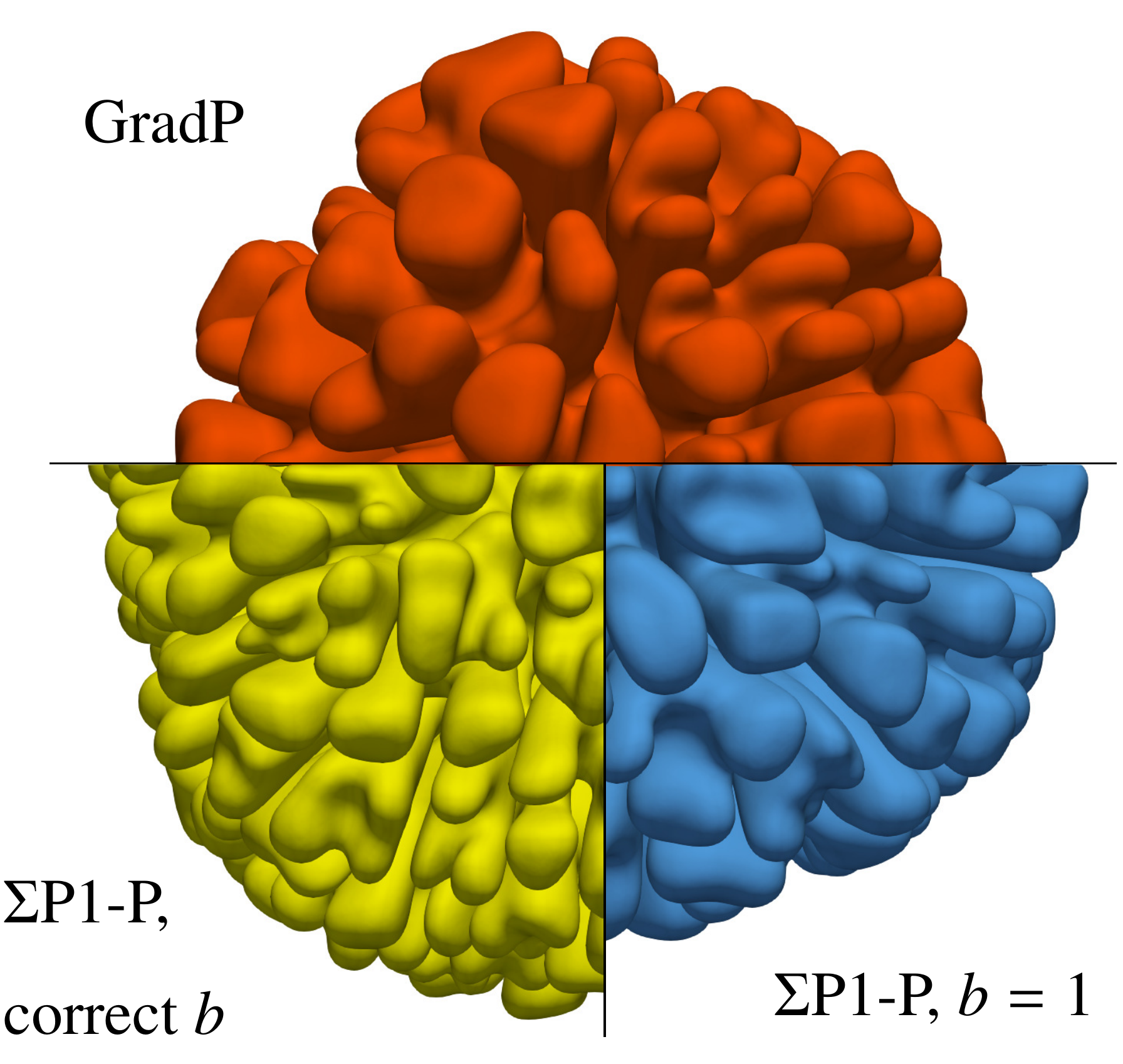}
\par\end{centering}
\caption{\label{fig:ISO-model-comparison}Comparison of crystal shapes obtained
by the respective reaction term models in isotropic crystal growth
simulations. Results without (left) and with (right) random perturbations
of the temperature field. Details on simulation setup are in Section
\ref{subsec:Isotropic-Simulations}.}
\end{figure}

Figure \ref{fig:ISO-model-comparison} demonstrates the solution of
the simulations of isotropic crystal growth at time $t=0.36$. The
settings are the same as in Figure \ref{fig:4-fold-model-comparison}
(see Section \ref{subsec:4-fold-Anisotropic-Simulations}) except
for the anisotropy strength $A_{1}=0$, which corresponds to using
the isotropic phase field equation (\ref{eq:Allen-Cahn-iso}) instead
of (\ref{eq:Allen-Cahn-anis}). In such case, the models GradP and
$\phi^{0}$GradP coincide and the $\Sigma$P1-P model has the form
such that the numerical analysis we performed in \cite{arXiv-PhaseField-FVM-Convergence}
applies. Results obtained both without ($\delta=0$) and with ($\delta=0.05$)
noise given by (\ref{eq:u-noise-field}) are provided at the same
scale, which testifies to the ability of noise to support crystal
growth and branching in a seaweed pattern.

\subsection{\label{subsec:Physically-Realistic-Simulations}Physically realistic
simulations}

Finally, we investigate the behavior of the individual models in the
simulation of rapid dendritic solidification of pure nickel. For this
situation, a range of both experimental \cite{Lum_Ni_rapid_solidification_experiment,Willnecker_Herlach-rapid_solidification_nonequilibrium,Herlach-Non-equilibrium-solidification-metals}
and computational \cite{Bragard_Karma-Higly_undercooled_solidif,Hoyt_Karma-Atomistic-continuum-modeling-solidif}
results is available in the literature. The models involved in the
study are $\phi^{0}$GradP, $\Sigma\phi^{0}$GradP, $\Sigma$P1-P
without the noise term and $\Sigma$P1-P with temperature field perturbation
described in Section \ref{subsec:Modeling-Irregular-Growth}. The
geometrical setup and initial and boundary conditions are described
in Section \ref{subsec:Basic-simulations-setup}. The dimensionless
parameters are set using the relations in Table \ref{tab:Dimensionless-quantities}
and the physical properties of Ni summarized in Table \ref{tab:Ni-solidification-parameters}.
Table \ref{tab:Ni-solidification-parameters} also contains the relevant
computational parameters. Again, assuming growth symmetry (see Section
\ref{subsec:Basic-simulations-setup}), the nucleation site is located
in the corner of $\Omega$ and its radius in dimensionless coordinates
is $0.02$. The 2nd order approximation of the anisotropic surface
energy for Ni is given by 
\begin{equation}
\begin{aligned}\psi\left(\vec{n}\right) & =1+A_{1}\left(n_{1}^{4}+n_{2}^{4}+n_{3}^{4}-\frac{3}{5}\right)\\
 & +A_{2}\left[3\left(n_{1}^{4}+n_{2}^{4}+n_{3}^{4}\right)+66n_{1}^{2}n_{2}^{2}n_{3}^{2}-\frac{17}{7}\right]
\end{aligned}
\label{eq:psi_6-fold_anisotropy}
\end{equation}
with $A_{1}=0.09,A_{2}=-0.011$ \cite{Hoyt_Karma-Atomistic-continuum-modeling-solidif}.
\begin{table}
\caption{\label{tab:Ni-solidification-parameters}Physical and computational
parameters of rapid solidification of pure Ni under very large supercooling.
The relation $b=1$ was used for the $\phi^{0}$GradP and $\Sigma\phi^{0}$GradP
models and both relations $b=1$ and (\ref{eq:b_COMPENSATION}) were
used for the $\Sigma$P1-P model.}

\centering{}%
\begin{tabular}{clcl}
\toprule 
Phys. qty & Value & Param. & Value\tabularnewline
\midrule
$\Delta u_{\text{ini}}$ & $50-300$ $\text{K}$ & $L_{0}$ & $10\text{ }$$\mu$$\text{m}$\tabularnewline
$\rho$ & $8900$ $\text{kg}\cdot\text{m}^{-3}$ & $\ell$ & $2$\tabularnewline
$c$ & $609$ $\text{J}\cdot\text{kg}^{-1}\text{K}^{-1}$ & $N$ & $480$\tabularnewline
$\lambda$ & $54.2$ $\text{W}\cdot\text{m}^{-1}\cdot\text{K}^{-1}$ & $\xi$ & $0.002$\tabularnewline
$L$ & $2.35\times10^{9}$ $\text{J}\cdot\text{m}^{-3}$ & $\varepsilon_{0}$ & $0.05$\tabularnewline
$u^{*}$ & $1728$ $\text{K}$ & $\varepsilon_{1}$ & $0.2$\tabularnewline
$\sigma$ & $0.37$ $\text{J}\cdot\text{m}^{-2}$ & $b$ & see caption\tabularnewline
$\Delta s$ & $1.36\times10^{6}$ $\text{J}\cdot\text{m}^{-3}\cdot\text{K}$ & $\delta$ & $0.05$\tabularnewline
$\mu$ & $1.99$ $\text{m}\cdot\text{s}^{-1}\cdot\text{K}^{-1}$ &  & \tabularnewline
\bottomrule
\end{tabular}
\end{table}

Besides observing the crystal morphology, the quantitative evaluation
of the simulations consists in calculating the dendrite tip velocity
as a function of initial supercooling $\Delta u_{\text{ini}}$ , which
is directly compared with the experimental results obtained in \cite{Willnecker_Herlach-rapid_solidification_nonequilibrium}
and \cite{Lum_Ni_rapid_solidification_experiment} and the phase-field
simulations performed in \cite{Bragard_Karma-Higly_undercooled_solidif,Hoyt_Karma-Atomistic-continuum-modeling-solidif}.

\subsubsection{Reasons for the choice of computational parameters}

A number of computational experiments with the values of $\Delta u_{\text{ini}}$
in the range given by Table \ref{tab:Ni-solidification-parameters}
showed that the correct formation of the diffuse but thin phase interface
requires the value of $\xi$ to be small enough, but still several
orders of magnitude larger than the interface thickness observed in
atomistic simulations \cite{Hoyt_Karma-Atomistic-continuum-modeling-solidif}.
This in turn leads to a minimum requirement on mesh resolution (see
Section \ref{subsec:Results-of-matched-asymptotics}). The length
scale $L_{0}$ and the size of the domain $\Omega$ given in Table
\ref{tab:Ni-solidification-parameters} correspond to the domain side
length in real coordinates $\ell L_{0}=20\text{ }\mu\text{m}$. For
such settings, the computational costs of the simulations remained
feasible: the wall times until the dendrite reached the opposite domain
boundary were between 24 and 48 hours on the HELIOS cluster at our
department. Each simulation employed three compute nodes equipped
with two 16-core AMD EPYC 7281@2.1GHz CPUs.

Dendrite tip length measurements were performed at solution snapshots
corresponding to physical times up to $2\text{ }\mu\text{s}$. Multiple
snapshots were used to ensure that the steady-state dendrite tip velocity
was calculated. In comparison, simulations in \cite{Bragard_Karma-Higly_undercooled_solidif}
(also with symmetry taken into account) had a supercooling-dependent
domain size up to $8\text{ }\mu\text{m}$ and the experiments in \cite{Bragard_Karma-Higly_undercooled_solidif},
\cite{Lum_Ni_rapid_solidification_experiment} were performed on a
time scale of tens of~$\mu\text{s}$.

The choice of the parameters $\varepsilon_{0},\varepsilon_{1}$ of
the $\Sigma$P1-P model was inspired by the findings of Section \ref{subsec:4-fold-Anisotropic-Simulations}.
In addition, numerical experiments revealed that only some of the
combinations investigated earlier in Section \ref{subsec:Choice-of-Eps}
could be used. For example, with $\varepsilon_{0}=0$ or $\varepsilon_{1}<0.15$,
the phase interface did not form correctly and spurious nucleation
sites appeared all over the domain $\Omega$.

\subsubsection{Results}

\begin{figure}
\begin{centering}
\includegraphics[width=0.95\figwidth]{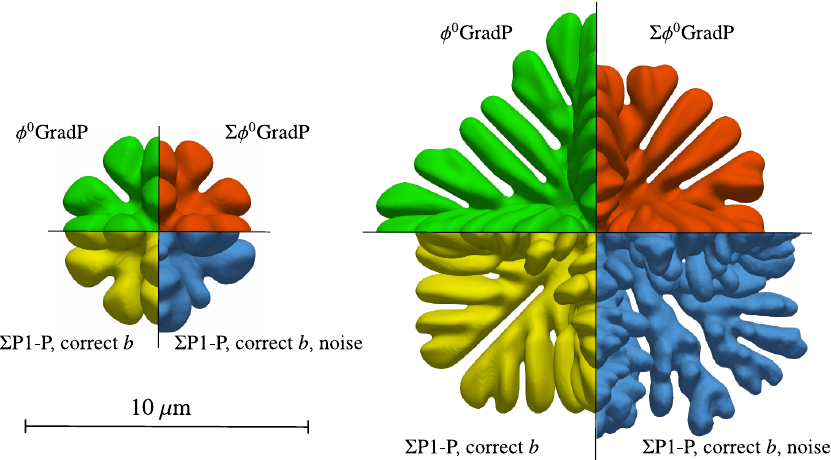}
\par\end{centering}
\caption{\label{fig:Ni-growth-deltauini-50-80K}Comparison of crystal shapes
obtained by different models for rapid Ni solidification at time $t=1.3\ \mu\text{s}$
and the initial supercooling $\Delta u_{\text{ini}}=50\text{ K}$
(left) and $\Delta u_{\text{ini}}=80\text{ K}$ (right).}

\end{figure}

With all four models (see the introduction to Section \ref{subsec:Physically-Realistic-Simulations}
above), simulations for initial supercooling values 
\[
\Delta u_{\text{ini}}\in\left\{ 50\text{ K},80\text{ K},120\text{ K},200\text{ K},300\text{ K}\right\} 
\]
 were performed. For $\Delta u_{\text{ini}}=50\text{ K}$, Figure
\ref{fig:Ni-growth-deltauini-50-80K} shows (using parallel projection
to the $xy$ plane) that the crystal morphology is similar for all
models. For $\Delta u_{\text{ini}}=80\text{ K}$, the original $\phi^{0}$GradP
model appears to prefer growth along coordinate axes, whereas the
remaining model variants produce the longest dendrite in the diagonal
direction (which is the expected behavior with anisotropy given by
(\ref{eq:psi_6-fold_anisotropy}) and $A_{1},A$ settings given in
Section \ref{subsec:Physically-Realistic-Simulations}). Comparing
the results of the $\Sigma$P1-P with and without noise, it seems
like noise promotes the dendrite growth, which was however not confirmed
as a general rule by the simulations with deeper supercooling.

\begin{figure}
\begin{centering}
\includegraphics[width=0.95\figwidth]{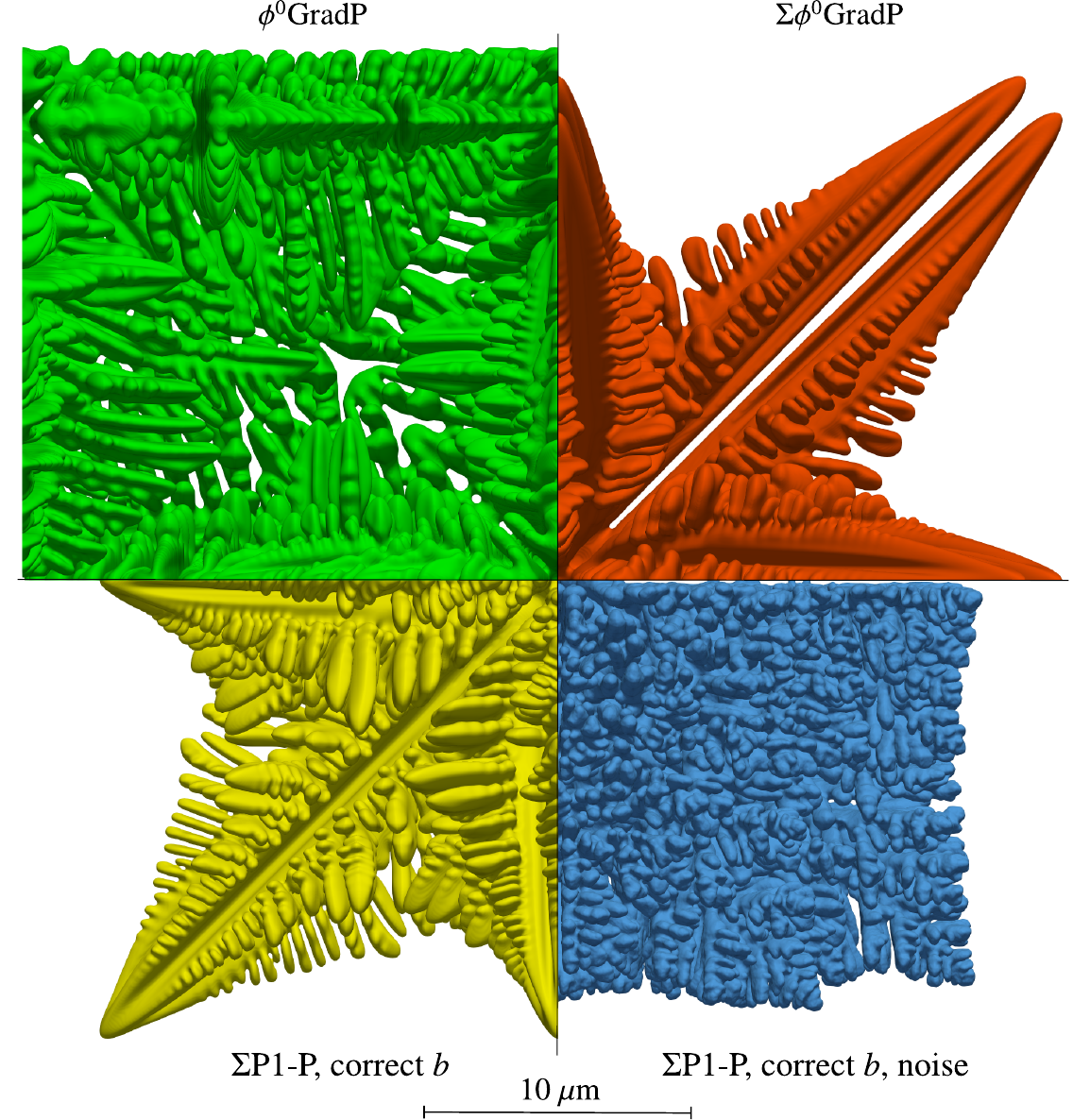}
\par\end{centering}
\caption{\label{fig:Ni-growth-deltauini-120K}Comparison of crystal shapes
obtained by different models for rapid Ni solidification at time $t=1.3\ \mu\text{s}$
and the initial supercooling $\Delta u_{\text{ini}}=120\text{ K}$.
The $\phi^{0}$GradP model fails as spurious nucleation sites occur.}
\end{figure}

With $\Delta u_{\text{ini}}=120\text{ K}$, $\phi^{0}$GradP fails
after some time, as can be seen in Figure \ref{fig:Ni-growth-deltauini-120K}.
A secondary nucleation site appears spontaneously and another crystal
grows from the opposite side of the domain. This is because $\phi^{0}$GradP
is not able to maintain the shape of the thin diffuse phase interface
and positive values of $p$ extend far away from the initial position
of the interface. The $\Sigma$-limited models ($\Sigma$$\phi^{0}$GradP
and $\Sigma$P1-P) remain stable and both produce qualitatively similar
crystal shapes. Dendrite tip twinning seen with the $\Sigma$$\phi^{0}$GradP
has been observed with the other models too during the numerical experiments
and the conditions for its emergence are worth further investigation
\cite{Kim_Takaki-Morphology_study}. As in Section \ref{subsec:4-fold-Anisotropic-Simulations},
Figure \ref{fig:Ni-growth-deltauini-120K} also shows that noise induces
side branching and the transition to a seaweed pattern, which becomes
even more pronounced as $\Delta u_{\text{ini}}$ increases further
up to $300\text{ K}$ (but this is not demonstrated visually here).
In general, our simulations have produced a much more complex and
developed dendritic morphology than those in \cite{Bragard_Karma-Higly_undercooled_solidif}.

\begin{figure}
\begin{centering}
\includegraphics[width=0.95\figwidth]{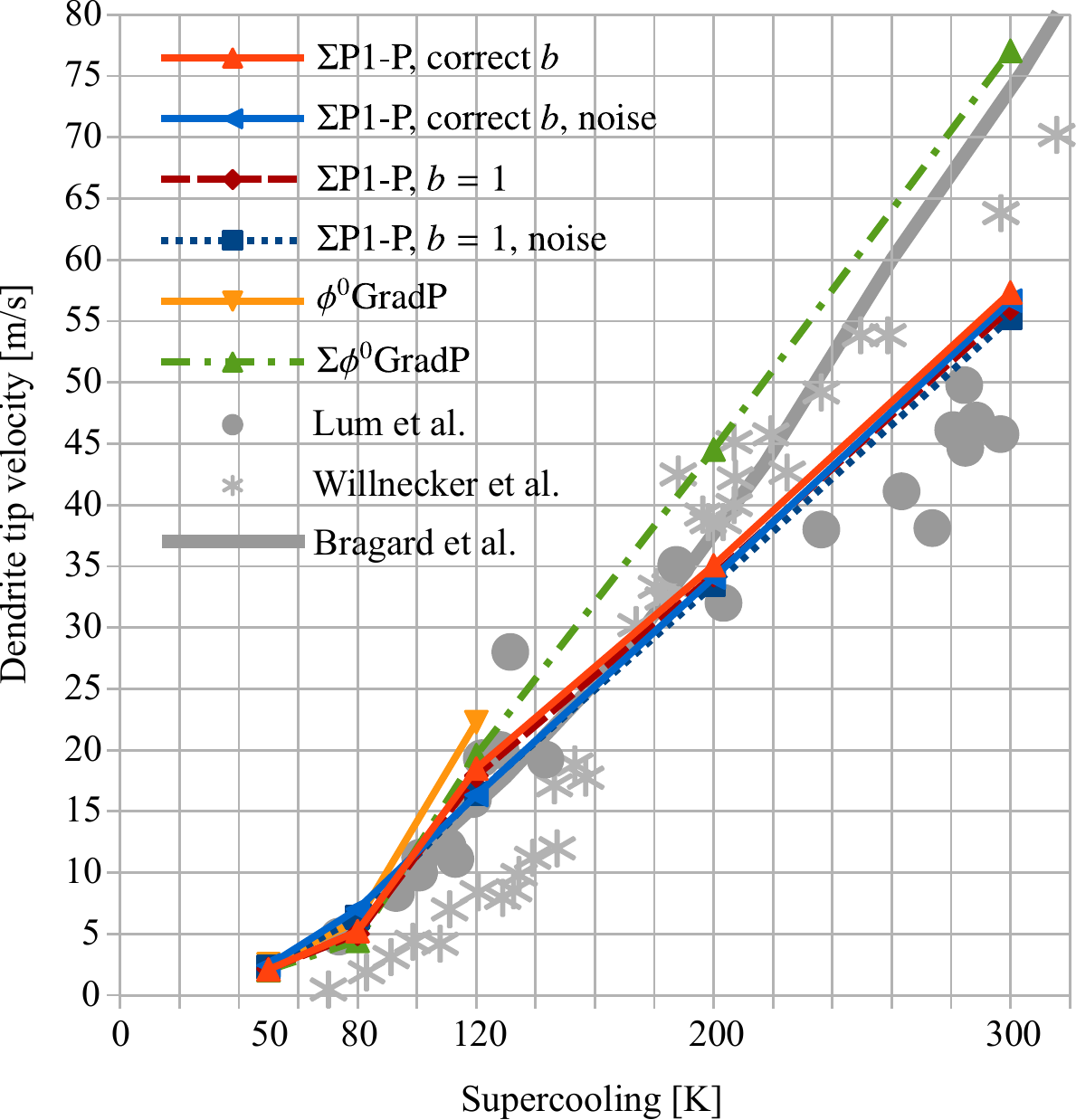}
\par\end{centering}
\caption{\label{fig:Ni-dendrite-tip-velocity}Dendrite tip velocity in pure
Ni solidification for large values of supercooling. Results obtained
by the different models compared to experimental data by Lum et al.
\cite{Lum_Ni_rapid_solidification_experiment} and Willnecker et al.
\cite{Willnecker_Herlach-rapid_solidification_nonequilibrium} and
to phase field simulations by Bragard et al. \cite{Bragard_Karma-Higly_undercooled_solidif}.}

\end{figure}

\begin{figure}
\begin{centering}
\includegraphics[width=0.95\figwidth]{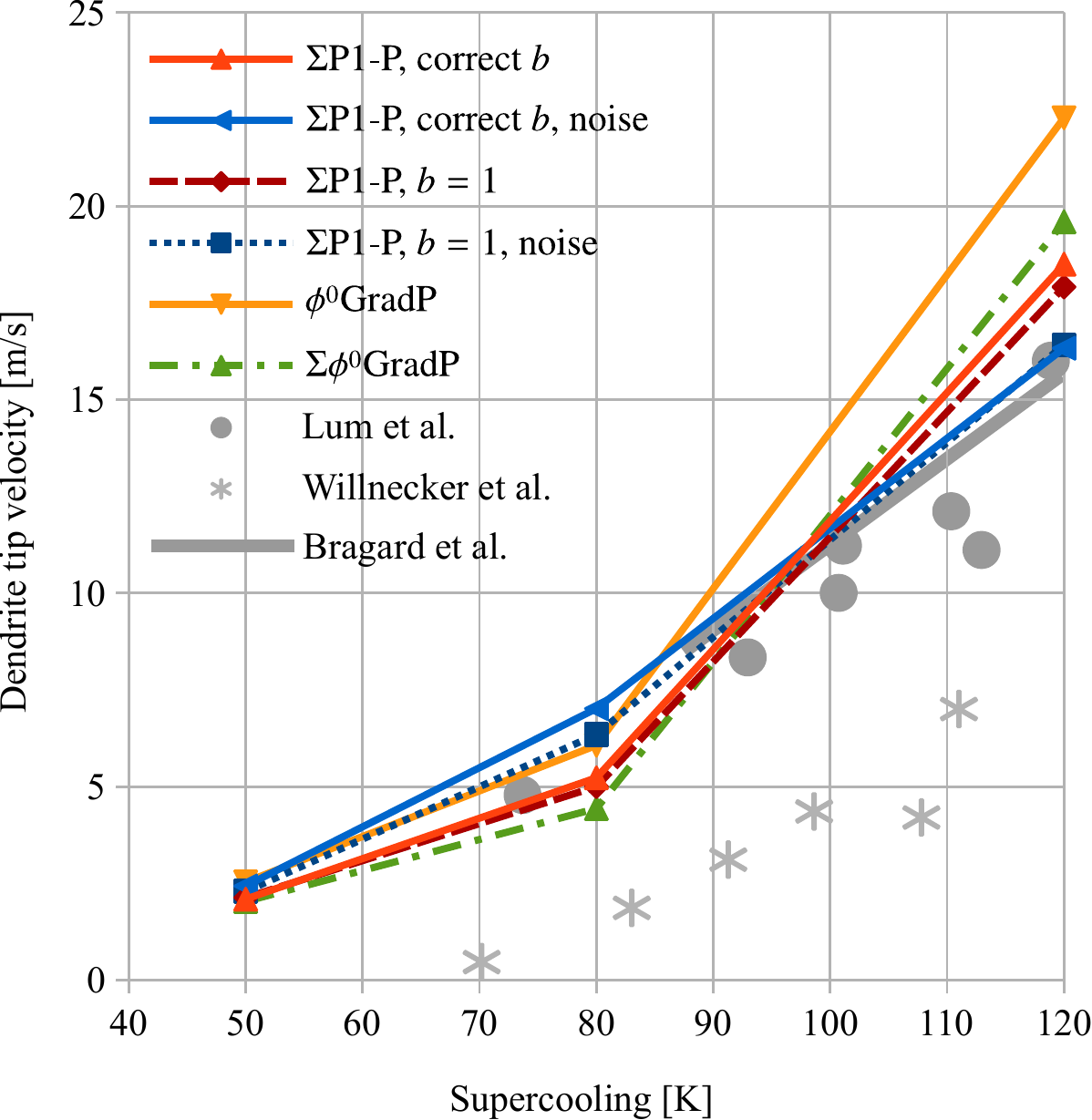}
\par\end{centering}
\caption{\label{fig:Ni-dendrite-tip-velocity-detail}Detail of the plot of
dendrite tip velocity in pure Ni solidification for supercooling up
to $120\text{ K}$. Results obtained by the different models compared
to experimental data by Lum et al. \cite{Lum_Ni_rapid_solidification_experiment}
and Willnecker et al. \cite{Willnecker_Herlach-rapid_solidification_nonequilibrium}
and to phase field simulations by Bragard et al. \cite{Bragard_Karma-Higly_undercooled_solidif}.}
\end{figure}

The dependence of dendrite tip velocity on $\Delta u_{\text{ini}}$
obtained by simulations with different models together with reference
data from literature are plotted in Figures \ref{fig:Ni-dendrite-tip-velocity}
and \ref{fig:Ni-dendrite-tip-velocity-detail}. The original $\phi^{0}$GradP
model failed for $\Delta u_{\text{ini}}\geq200$ with spurious nucleation
sites spreading all over the computational domain almost immediately
after starting the simulation despite $\Delta u_{\text{ini}}$ being
far from the expected onset of homogeneous nucleation \cite{Filipponi-Nucleation_in_undercooled_Nickel}.
The other models exhibit a reasonable agreement with the reference
data. The $\Sigma$$\phi^{0}$GradP model gives the highest predictions
of dendrite tip velocity slightly above the computations by Bragard
et al. \cite{Bragard_Karma-Higly_undercooled_solidif}. The $\Sigma$P1-P
model in all variants yields very similar results that best agree
with the measurements by Lum et al. \cite{Lum_Ni_rapid_solidification_experiment},
indicating the transition from power-law to linear dependence described
both by Lum et al. \cite{Lum_Ni_rapid_solidification_experiment}
and Willnecker et al. \cite{Willnecker_Herlach-rapid_solidification_nonequilibrium}.
The setting of $b$ has a negligible effect. Adding noise does not
generally increase the tip velocity, but it makes the curve more straight
(cf. Figure \ref{fig:Ni-growth-deltauini-50-80K} and the related
discussion).

\section{Conclusion}

We recalled the asymptotic correspondence \cite{Caginalp-Stefan_HeleShaw_PF,Benes-Asymptotics}
between the phase field models used in our previous computational
studies \cite{ENUMATH2011,ISPMA14-Pavel_Ales-ActaPhPoloA} and the
sharp interface formulation \cite{Gurtin-Stefan_problem}. It is possible
to design reaction terms implying different distributions of latent
heat release rate across the diffuse interface while keeping the asymptotic
properties of the resulting phase field model valid. Inspired by the
computationally favorable behavior of the GradP and $\phi^{0}$GradP
models, we proposed an alternative form of the reaction term with
a similar latent heat distribution, but completely avoiding the gradient
term in (\ref{eq:Gradient-model}). The resulting term required further
treatment by the $\Sigma$ limiter (\ref{eq:Sigma-limiter}) to maintain
the shape of the thin diffuse interface for larger values of the initial
supercooling $\Delta u_{\text{ini}}$, which gave rise to the $\Sigma$P1-P
model. Unlike the GradP model, the $\Sigma$P1-P model in its isotropic
form is compatible with the numerical analysis performed in our related
work \cite{arXiv-PhaseField-FVM-Convergence}, which provides theoretical
justification of the proper function of the implemented finite volume-based
numerical solvers. However, the $\Sigma$ limiter can also be applied
to the original GradP and $\phi^{0}$GradP models.

The results of the computations the $\Sigma$P1-P model are sensitive
to the settings of the $\Sigma$ limiter parameters $\varepsilon_{0},\varepsilon_{1}$.
By numerical experiments, we found a setting that produces very similar
crystal morphologies to the results of the GradP and $\phi^{0}$GradP
models. This setting was also used for successful simulations of rapid
solidification of pure nickel that were not possible at all with the
GradP and $\phi^{0}$GradP models. However, these promising results
deserve additional computational studies. Finer sampling of the initial
supercooling range should confirm the model's capability of reproducing
the transition from power-law to linear character of the curve in
Figure \ref{fig:Ni-dendrite-tip-velocity}, which is observed in the
experimental data \cite{Willnecker_Herlach-rapid_solidification_nonequilibrium,Lum_Ni_rapid_solidification_experiment}
but not predicted by the earlier phase field simulations \cite{Bragard_Karma-Higly_undercooled_solidif}.
Simulations with varying anisotropy strength parameters $A_{1},A_{2}$
should reveal the behavior of the proposed model with respect to the
resulting crystal morphology, which can be compared e.g. with the
recent results in \cite{Kim_Takaki-Morphology_study}. In the future,
access to more powerful computational resources may allow us to study
the models while further decreasing the thickness of the diffuse interface
controlled by $\xi$.

\noindent 

\section*{Data availability}

The datasets and computer codes are available upon request from the
authors.

\section*{Declaration of Competing Interest}

The authors declare that they have no known competing financial interests
or personal relationships that could have appeared to influence the
work reported in this paper.

\section*{CRediT authorship contribution statement}

\textbf{Pavel Strachota:} Conceptualization, Methodology, Software,
Validation, Formal analysis, Investigation, Data Curation, Writing
- Original Draft, Writing - Review \& Editing, Visualization. \textbf{Aleš
Wodecki:} Methodology, Writing - Review \& Editing. \textbf{Michal
Beneš:} Writing - Review \& Editing, Supervision, Project administration.

\section*{Acknowledgments}

This work is part of the project \emph{Centre of Advanced Applied
Sciences} (Reg. No. CZ.02.1.01/0.0/0.0/16-019/0000778), co-financed
by the European Union. Partial support of grant No. SGS20/184/OHK4/3T/14
of the Grant Agency of the Czech Technical University in Prague.



\bibliographystyle{model1-num-names}
\bibliography{References/publications,References/references_MR-DTI,References/references_MATH-PHYS,References/references_IT}

\end{document}